\documentclass[12pt,a4paper]{article}

\usepackage[T1]{fontenc}
\usepackage[hmarginratio=1:1,top=32mm,columnsep=20pt]{geometry}
\usepackage{mathdots}
\usepackage{mathrsfs}
\usepackage{color}
\definecolor{dark-gray}{gray}{0.20}
\definecolor{gray}{gray}{0.30}
\definecolor{light-gray}{gray}{0.80}
\definecolor{dark-red}{rgb}{0.7,0,0}
\definecolor{dark-green}{rgb}{0.1,0.4,0}
\definecolor{dark-blue}{rgb}{0.3,0.3,0.7}
\definecolor{light-blue}{rgb}{0.8,0.8,1}

\usepackage{cite}
\usepackage{hyperref}
\hypersetup{
	colorlinks=true,
	linkcolor=dark-blue,
	citecolor=dark-red,
	urlcolor=dark-green,
	linktoc=page
}

  \usepackage{a4wide}
  \usepackage{latexsym}
  \usepackage{epsf}
  \usepackage{bbm}
  \usepackage{graphicx}
  \usepackage{amsmath,amssymb,amsthm}
  \usepackage{verbatim}

\newcommand{\e}{\mathrm{e}}

\newcommand{\bbm}{\left(\begin{matrix}}
\newcommand{\ebm}{\end{matrix}\right)}
\newcommand{\bea}{\begin{eqnarray}}
\newcommand{\eea}{\end{eqnarray}}
\newcommand{\be}{\begin{equation}}
\newcommand{\ee}{\end{equation}}

\renewcommand{\d}{\textrm{d}}

\newcommand{\SO}{\mathop{\rm SO}}

\begin{document}

\numberwithin{equation}{section}

\begin{center}

{\LARGE {\bf Instantons and (no) wormholes in $AdS_3\times S^3 \times CY_2$ }}  \\

\vspace{1.5 cm} {\large  D. Astesiano$^a$, D. Ruggeri $^b$, M. Trigiante$^b$ and  T. Van Riet$^{c}$ }\footnote{{ \upshape\ttfamily dastesiano@uninsubria.it, daniele.rug@gmail.com, mario.trigiante@polito.it,  thomas.vanriet @fys.kuleuven.be
 } }\\
\vspace{1 cm}  ${}^a$ Department of Science and High Technology, Universita dell'Insubria,\\
Via Valleggio 11, 22100, Como, Italy
$\&$ INFN, sezione di Milano\\ Via Celoria 16, 20133, Milano, Italy
\\ \vspace{.15 cm}
\vspace{.15 cm} { ${}^b$Department of Applied Science and Technology, Politecnico di Torino; \\ C.so Duca degli Abruzzi, 24, I-10129 Torino, Italy,\\ 
INFN–Sezione di Torino, Via P. Giuria 1, 10125 Torino, It\\\vspace{.15 cm}
${}^c$Instituut voor Theoretische Fysica, K.U. Leuven,\\
Celestijnenlaan 200D B-3001 Leuven, Belgium }

\vspace{2cm}

{\bf Abstract}
\end{center}

{\small We study supergravity instantons sourced by axion (and saxion) fields in the Euclidean $AdS_3\times S^3 \times CY_2$ vacua of IIB supergravity. Such instantons are described by geodesic curves on the moduli space; the timelike geodesics can describe Euclidean wormholes, the lightlike geodesics describe (generalisations of) D-instantons and spacelike geodesics are sub-extremal versions thereof. We perform a concrete classification of such geodesics and find that, despite earlier claims, the wormholes fail to be regular. A subclass of the lightlike geodesics is supersymmetric and, up to dualities, lift to Euclidean strings wrapping 2-cycles in the CY$_2$. The dual of these instantons are expected to be worldsheet instantons of the D1-D5 CFT.  

\setcounter{tocdepth}{2}
\newpage
\textcolor{white}{x}\\

\vspace{6cm}
\begin{center}
    \emph{In memory of Mees de Roo and Luciano Girardello}
\end{center}
\newpage
\tableofcontents
\newpage

\section{Introduction}
The study of the Euclidean path integral for gravity has a long history with recent breakthroughs for low-dimensional gravity theories like JT gravity, see eg \cite{Marolf:2020xie}. One of the main lessons to be learned from  low-dimensional theories is that it does make sense to sum over saddle points with different topologies and they tend to have a holographic description in terms of ensemble theories, akin to Coleman's $\alpha$-parameters and the associated absorbtion and emission of baby universes \cite{Coleman:1988cy}. However the rules of the game for actual Einstein-Hilbert gravity (coupled to matter) in dimensions 3 and higher remains somewhat unclear. It has even been suggested that it is vastly different from the lessons learned in low-dimensional gravity theories. The role Euclidean wormholes\footnote{Reviewed nicely in \cite{Hebecker:2018ofv}.} can play is not clear picture at this point in time. In fact some well-motivated ideas on quantum gravity and the Swampland \cite{VanRiet:2020pcn, McNamara:2020uza} suggest that wormholes do not contribute in the ways envisaged in the early works \cite{Coleman:1988cy, Lavrelashvili:1987jg, Giddings:1987cg} (see also \cite{Heckman:2021vzx}).

String theory provides a UV completion of quantum gravity and therefore various ideas on the semi-classical formulation of gravity should be testable. Since saddle point expansions are non-perturbative, one is naturally led to consider holographic dual pairs in string theory as they provide a non-perturbative definition of string theory in certain AdS backgrounds.  This topic was initiated in some early works \cite{Maldacena:2004rf, Bergshoeff:2005zf, ArkaniHamed:2007js} (see also \cite{Betzios:2019rds, Betzios:2021fnm, Kundu:2021nwp})
 with the hope that AdS/CFT should inform us about which saddle points contribute and how. In this regard the most natural Euclidean wormholes to consider are wormholes sourced by axion fields \cite{Giddings:1987cg}, as axions provide a natural source of negative Euclidean energy-momentum required to sustain a wormhole geometry. As emphasized in \cite{Gutperle:2002km, Bergshoeff:2004pg, Bergshoeff:2005zf,VanRiet:2020csu, VanRiet:2020pcn} such set-ups come with a bonus: axions in string theory pair up with saxions who have positive energy-momentum instead and one can find configurations which interpolate between negative (wormholes) through zero, towards positive (Euclidean) energy-momentum (EM). An example of the zero EM solutions are D(-1) branes, aka D-instantons \cite{Gibbons:1995vg, Green:1997tv}. They can be supersymmetric (SUSY) and their role in string theory is well understood. 
 Similar to black holes and branes one can think in terms of an extremality property:
\begin{itemize}
	\item Negative energy-momentum $\rightarrow$ wormholes as super-extremal instantons
	\item Zero energy-momentum $\rightarrow$ extremal instantons
	\item Positive energy-momentum $\rightarrow$ sub-extremal solutions.
\end{itemize}
This picture can be made explicit by the c-map (reduction over time) where the extremality properties of black holes reduce to the corresponding extremality properties of instantons \cite{Breitenlohner:1987dg, Bergshoeff:2004pg, Bergshoeff:2008be, Bossard:2009at,  Mohaupt:2011aa}. Alternative (but related) viewpoints are obtained by computing the action to charge ratio \cite{Montero:2015ofa, Brown:2015iha, Hebecker:2016dsw} or using probe D-instanton actions to infer Euclidean ``repulsion or attraction'' \cite{VanRiet:2020pcn, VanRiet:2020csu}.

There is however one caveat with the c-map picture: the wormholes obtained from reducing super-extremal black holes have badly singular axion-saxion profiles which lift to the naked singularity of the super-extremal ``black hole''. To obtain smooth wormholes one needs a specific inequality to hold on the axion-saxion coupling \cite{ArkaniHamed:2007js} which was claimed to be possible in Euclidean $AdS_3\times S^3 \times CY_2$ \cite{ArkaniHamed:2007js} and found more recently in Euclidean $AdS_5\times S^5/Z_k$ with $k>1$ \cite{Hertog:2017owm}. One of the results of this paper is that the regularity condition is in fact not satisfied in $AdS_3\times S^3 \times CY_2$, making the 5D examples of \cite{Hertog:2017owm} the sofar unique embeddings of axion wormholes in AdS.   

Using the latter explicit embedding and its dual $N=2$ necklace quiver description \cite{Kachru:1998ys, Corrado:2002wx, Louis:2015dca} some properties of the instantons could be inferred in \cite{Ruggeri:2017grz, Katmadas:2018ksp}: the extremal instantons where argued to map to specific SUSY and non-SUSY instantons of the gauge theory. The SUSY instantons have the same orientation of the (anti-) self-dual gauge fields at every gauge node, whereas the non-SUSY instantons had at least one gauge node with opposite orientation. The sub-extremal solutions remain unclear and a speculative description in terms of non-self dual gauge field configurations was given earlier in \cite{Bergshoeff:2005zf} for $AdS_5 \times S^5$ and is readily extended to $AdS_5\times S^5/Z_k$. The super-extremal solutions (the wormholes) are problematic since the holographic one-point functions violate a positivity bound suggesting the wormholes are in fact unphysical. We interpret this as a manifestation of the recently discovered infinite number of perturbative instabilities (negative modes) of 4D axion wormholes sourced by a single axion \cite{Hertog:2017owm}, correcting earlier contradicting claims in \cite{Rubakov:1996cn, Alonso:2017avz}. It is natural to expect that the instabilities also arise when multiple axions and saxions interact in some general sigma model \cite{Progress}. If so, the macroscopic wormholes cannot contribute to the path integral whereas a similar configuration of widely separated microscopically sized solutions with unit axion charge, should be the dominant saddle points \cite{VanRiet:2020csu, VanRiet:2020pcn}\footnote{Interesting recent work has reported on ``brane-nucleation" type instabilities (different from negative modes) in general classes of Euclidean AdS wormholes embedded in string/M theory \cite{Marolf:2021kjc}.
}. They are however outside of the supergravity regime and should not be interpreted as wormhole geometries.

In this paper we continue our investigation of AdS moduli spaces, their geodesics and the relation with supergravity and CFT instantons initiated in \cite{Hertog:2017owm, Ruggeri:2017grz, Katmadas:2018ksp} and extend it to Euclidean $AdS_3\times S^3 \times CY_2$ with $CY_2$ either $\mathbb{T}^4$ or $K_3$. This holographic background is well-studied and its dual CFT, known as the D1-D5 CFT, has a Lagrangian description in the free orbifold limit \cite{Giveon:1998ns, Avery:2010qw, David:2002wn}.  Despite $AdS_3\times S^3 \times CY_2$ being one of the most well-known AdS/CFT backgrounds there has been surprisingly little investigation of the instantons in these backgrounds up to two works we are aware of \cite{ArkaniHamed:2007js, Kogan:1998re}. This is in rather stark contrast with the study of instantons in $AdS_5\times S^5$ \cite{Banks:1998nr, Bianchi:1998nk, Dorey:1998xe, Belitsky:2000ws, Vandoren:2008xg} which constitutes one of the main early breakthroughs in our understanding of AdS/CFT.  The aim of this paper is to carefully classify the instantons with O(3) symmetry sourced by the AdS moduli (axions and saxions), which boils down to solve for and classify geodesic curves on the moduli space. We will find disagreement with the earlier investigations of \cite{ArkaniHamed:2007js, Kogan:1998re}. A dual description of the extremal supersymmetric instantons in terms of instantons in the D1-D5 CFT is left for a follow-up work. 

\section{The general set-up}
The strategy of \cite{Hertog:2017owm, Ruggeri:2017grz,Katmadas:2018ksp} to embed Euclidean axion wormholes in AdS compactifications of 10d/11d supergravity is to truncate the compactification down to its moduli space of scalars such that the resulting Lagrangian after the truncation reads:
\be\label{generalaction}
e^{-1}\mathcal{L} = \mathcal{R} -\tfrac{1}{2} \mathscr{G}_{IJ}\partial \phi^I\partial\phi_I - \Lambda\,,
\ee
where the $\phi^I$ are the AdS moduli, $\mathscr{G}_{IJ}$ the metric on moduli space and $\Lambda$ the negative vacuum energy at the AdS critical point of the scalar potential. In this definition moduli are not just massless, but they have no appearance whatsoever in the effective potential at the vacuum\footnote{So no cubic or higher couplings.}. A holographic dual statement is that the dual operators are \emph{exactly} marginal and the moduli space is then dual to the conformal manifold describing a (continous) set of CFTs labeled by the vevs of the moduli dual to the values of the coefficients in front of the exactly marginal operators in the CFT Lagrangian (if any).

In Euclidean supergravity the moduli space metric is not necessarily positive definite. Even more, it seems that its signature is not uniquely fixed by supersymmetry since there are sign ambiguities in defining Euclidean (10D) sugra \cite{Bergshoeff:2007cg}. However this ambiguity is resolved if one wishes to study instantons sourced by axions since these instantons are interpreted as axion charge transitions and then the boundary conditions in the path integral fix completely the sign, see eg \cite{Burgess:1989da, Bergshoeff:2005zf}: axions get flipped signs (consistent with them enjoying a shift symmetry) and other scalars remain.

To make this paper self-contained we briefly review the general form of instanton solutions in $D>2$ as presented in detail in \cite{Breitenlohner:1987dg, Bergshoeff:2008be, VanRiet:2020pcn}. Once a radially symmetric instanton Ansatz is made:
\be
ds^2 = f(\tau)^2 \d\tau^2 + a(\tau)^2\d\Omega_{D-1}^2\,, \qquad \phi^I=\phi^I(\tau)\,,
\ee
the scalar field equations of motion are purely geodesic, and with the gauge choice, $f=a^{D-1}$ the geodesics have an affine parametrisation in terms of the harmonic function $\rho$ on the Euclidean geometry, consequently the geodesic velocity is a constant $c$:
\be
\mathscr{G}_{IJ}\partial_{\rho} \phi^I\partial_{\rho}\phi^J = c\,.
\ee
As emphasized earlier, due to the presence of axions, $c$ can have any sign and the solution for the metric is independent of the sigma model details. A particularly simple expression exist for the gauge $a=\tau$ can be found:
\be
f(\tau)^2 =\left( 1+ \frac{\tau^2}{\ell^2} + \frac{c}{2(D-2)(D-1)}\tau^{-2(D-2)} \right)^{-1}\,,
\ee
where $\Lambda=-(D-1)(D-2)/\ell^2$. For $c=0$ this is the metric on Euclidean AdS, for $c>0$ this is a singular solution whose metric coincides with the Euclidean version of Gubser's ``dilaton-driven confinement''-solution \cite{Gubser:1999pk} and for $c<0$ the metric describes a wormhole.\footnote{Note that these particular coordinates only cover half the wormhole since there is a  coordinate singularity at $\tau=\tau^*$ where $f(\tau^*)=\infty$ and there one can consistently glue a mirror copy to have the whole smooth wormhole metric. Other coordinates can make this more explicit but are not needed here and described in the references quoted earlier.} The wormhole metric $c<0$ is smooth in proper coordinates but the scalars do not need to be. There can be unphysical kink-like singularities as for instance observed for $AdS_5\times S^5$ \cite{Bergshoeff:2005zf}. A regularity condition, involving the length of timelike geodesics, was found in \cite{ArkaniHamed:2007js} and shown to be possible in $AdS_5\times S^5/\mathbb{Z}_k$ when $k>1$ \cite{Hertog:2017owm}.  In what follows we investigate the wormhole regularity criterion for $AdS_3\times S^3 \times CY_2$ in Euclidean IIB supergravity, and further construct all the solutions and check whether supersymmetry can be preserved for  $c=0$.

To answer the questions laid out above, we need to know the moduli space at stake. Early work by Cecotti \cite{Cecotti:1990kz} on 2D CFTs of the kind we expect to find, suggest conformal manifolds for the Lorentzian CFT of the form
\be
\frac{SO(4,n)}{SO(4)\times SO(n)}\,.\label{themanifold}
\ee
This is confirmed by AdS/CFT since the moduli spaces of  $\rm AdS_3\times S^3\times \mathbb{T}_4$ should be the one with $n=5$ \cite{Lu:1997bg, Andrianopoli:2007kz} and the moduli space of  $\rm AdS_3\times S^3\times K_3$ should have $n=20$ \cite{Aspinwall:1996mn}. Below we will construct the Wick-rotated version of this moduli space. But before we set up this general machinery of coset spaces and geodesic curves we take a different, more 10-dimensional viewpoint to find some simple truncations of the moduli space and their corresponding solutions. The group theory in Section \ref{GTSection} will then prove that the results obtained from this particularly simple truncation is extended to the whole moduli space. In other words, the truncation of the moduli space discussed in the next section contains the \emph{seed} solutions that generate all other solutions of interest which lie within a general set of orbits with respect to the isometry (duality) group of the full moduli space. In the same section we construct the generating solutions of all the geodesics in the moduli space. \par
The paper is organized as follows:\\
In Section \ref{sec:simple} we restrict ourselves to a simple truncation of the moduli space and derive therein the geodesics which describe the supersymmetric configuration corresponding to two Euclidean D1 branes in the D1-D5 background. The uplift of these geodescis to $D=10$ is performed and their on-shell action computed. We also prove that this simple truncation contains no regular wormhole solution.\par In Section \ref{GTSection}, using the theory of Lie groups and Lie algebras, we discuss how general the results obtained in the previous Section are. We rigorously define the Wick-rotation and prove that the duality orbits of geodesics in the moduli space, both in the $\mathbb{T}_4$ and in the $K_3$ cases, have a representative in the smaller Wick-rotated version of (\ref{themanifold}) with $n=4$. The classification of the extremal solutions will require tools borrowed from the theory of nilpotent orbits in classical Lie algebras. We shall prove, in Subsection \ref{normal}, that the Euclidean D1-brane solutions derived in Section \ref{sec:simple} are generating all order-2 nilpotent orbits (with reference to a suitable representation of $\mathfrak{so}(4,4)$). As for the remaining orbits, we give the explicit form of the generating extremal and non-extremal geodesics, in terms of the $D=10$ fields, in Subsection \ref{totalsolution}.  Finally, in Subsection \ref{remark}, we elaborate on the existence of regular wormholes and argue that, in light of the criterion put forward in \cite{ArkaniHamed:2007js}, the negative result of Subsection \ref{nowormholes} extends to the whole moduli space.\par We end with concluding remarks.

\section{A simple truncation and its solutions}\label{sec:simple}
\subsection{Brane intersections}
It is insightful to recall the brane-intersection whose near horizon gives the vacuum. The 10D brane picture is given by
 \begin{align}
  D1 & \times \times - - - - - - - - \nonumber \\
  D5 & \times \times- - - - \times \times\times\times \nonumber 
 \end{align}
 Our notation is such that, upon taking the near horizon limit, the first 3 directions generate $AdS_3$, then $S^3$ and then $\mathbb{T}^4$ or $K_3$. This picture naturally suggest the existence of SUSY instantons localised in $AdS_3$ from Euclidean D1 string wrapping two-cycles in the $CY_2$. In particular for $CY_2=\mathbb{T}^4$ we would have
 \begin{align}
D1 & \times \times - - - - - - - - \nonumber \\
D5 & \times \times - - - -\times \times\times\times \nonumber \\
D1 & - - - - - - - - \times \times \nonumber \\
D1 & - - - - - - \times \times - -\label{intersection}
\end{align}
The naive counting of supercharges works as follows: the D1-D5 intersection preserves 8 supercharges but doubles to 16 upon taking the near horizon limit. If only one stack of Euclidean D1 branes is present then SUSY is broken to 8 supercharges. The presence of both stacks would further reduce it to a configuration preserving 4 of the original 32.   An intersection diagram that contains a Euclidean D3 wrapping the $\mathbb{T}^4$ suggest that this preserve no SUSY (or even be a solution).

We will present a detailed analysis of the moduli space later in this paper, but we can already make some educated guesses as to where the potential axions in 3D can come from; integrating the NSNS and RR two-forms $B_2, C_2$ over the two-cycles in $CY_2$, integrating $C_4$ over the whole $CY_2$ and then the RR axion  $C_0$ itself.   Note that the KK vectors from $\mathbb{T}^4$, the vectors from $B_2, C_2$ over the $\mathbb{T}^4$ 1-cycles could be dualised to axions in 3D, but we should not. They are ``true vectors" and confine in 3D. This is related to the choice of boundary conditions for vectors in $AdS_3$ as explained in \cite{Montero:2016tif}.

The full moduli space from the $\mathbb{T}^4$ reduction will deliver too many axions, because some will be lifted by the $F_3$ flux on $AdS_3$ and $S^3$; one can demonstrate \cite{Giveon:1998ns, David:2002wn} that only 5 out of  the 25 $\mathbb{T}^4$ moduli\footnote{ 10 from the metric, 6 from the $B_2$ field, 6 from the $C_2$ field, the string coupling, one from the $C_4$ and one from $C_0$.} in 6D get lifted by the fluxes. These are the a linear combination of torus volume and dilaton, a linear combination of $C_0$ and $C_4$, and 3 from the $C_2$ field (the self dual ones). The remaining 20 moduli span the manifold
\be
\frac{\SO(5,4)}{\SO(4)\times\SO(5)}\,.
\ee

\subsection{Dimensional reduction and truncation}
The above discussion inspires to find a simple consistent truncation for $\mathbb{T}^4$. We keep the torus volume, the volume of a 2-cycle (which determines the volume of the orthogonal 2-cycle). Then we also keep 2 axions from $C_2$ reduced over these two 2-cycles and call them $c_1$ and $c_2$. The Ansatz in 10D Einstein frame is
\begin{align}
& ds^2_{10} = e^{2\alpha\varphi} ds^2_{6} + e^{2\beta\varphi}\left(e^{2\gamma\psi} d \theta_1^2 + e^{2\gamma\psi} d \theta_2^2+ e^{-2\gamma\psi} d \theta_3^2 + e^{-2\gamma\psi} d \theta_4^2\right)\,,\\
& \hat{C}_2 = C_2+  c_1 d\theta_1\wedge d\theta_2 + c_2 d\theta_3\wedge d\theta_4\,,
\end{align}
where $\alpha=1/4=-\beta$ and $\gamma^2=1/8$. The Lagrangian of our truncation in six dimensions is:
\begin{align}
e^{-1}\mathcal{L}=& \mathcal{R}_6 -\tfrac{1}{2}(\partial \phi)^2-\tfrac{1}{2}(\partial \varphi)^2-\tfrac{1}{2}(\partial \psi)^2 \nonumber\\ &-\tfrac{1}{2}e^{\phi+\varphi-4\gamma\psi}(\partial c_1)^2 -\tfrac{1}{2}e^{\phi+\varphi+4\gamma\psi}(\partial c_2)^2 - \tfrac{1}{2 3!}e^{\tfrac{1}{2}\varphi +\phi}F_3^2 \,.
\end{align}
Now we reduce further down to three dimensions using electric and magnetic flux
\begin{align}
&ds^2_6 = \e^{2\bar{\alpha}\bar{\varphi}} ds^2_3 + \e^{2\bar{\beta}\bar{\varphi}} \d\Omega_3^2\,,\\
&F_3=  Q_1 e^{-\phi} e^{3(\bar{\alpha}-\bar{\beta})\bar{\varphi}}\epsilon_3 + Q_5 \tilde{\epsilon}_3  \,,
\end{align}
where $\bar{\alpha}^2=3/8$, $\bar{\alpha}=-3\bar{\beta}$ and $\epsilon_3$ resp. and $\tilde{\epsilon}_3$ are the volume elements of $ds^2_3$ resp $\d\Omega_3^2$ and we introduced $\bar{\varphi}$, the volume scalar of the $S^3$. We then find\footnote{We normalise the curvature of the 3-sphere metric $\d\Omega_3^2$ to be six.}
\begin{equation}
e^{-1}\mathcal{L} = \mathcal{R}_3 - \text{kinetic}   - V(\phi,\psi,\varphi,\bar{\varphi}) \,,
\end{equation}
where
\begin{align}
&2\times\text{kinetic}  = (\partial \phi)^2 +(\partial \varphi)^2+(\partial \psi)^2 +(\partial \bar{\varphi})^2  +e^{\phi+\varphi}\left(e^{-4\gamma\psi}(\partial c_1)^2 +e^{4\gamma\psi}(\partial c_2)^2\right)\,, \\
 &V(\phi,\psi,\varphi,\bar{\varphi}) =  \tfrac{1}{2}Q_1^2e^{-(\phi-\varphi) + 4\bar{\alpha}\bar{\varphi}} + \tfrac{1}{2}Q_5^2 e^{(\phi-\varphi) + 4\bar{\alpha}\bar{\varphi}} - 6e^{-8\bar{\beta}\bar{\varphi}}\,.
\end{align}
Note that the dependence of the axion kinetic term on $\bar{\varphi}$ cancelled out.  The potential stabilises the scalar
$\bar{\varphi}$ and the combination $(\phi-\varphi)$. Interestingly, it is exactly the orthogonal combination $(\phi+\varphi)$ that is appearing in the axion kinetic term, which will proof necessary for our truncation. 
So, let us call
\be
\sqrt{2}\tilde{\phi} = \phi+\varphi\,.
\ee
We find that the 3D action, in the vacuum, truncates to:
\begin{align}
e^{-1}\mathcal{L} = R &-\tfrac{1}{2}(\partial \tilde{\phi})^2-\tfrac{1}{2}(\partial \psi)^2  -\tfrac{1}{2}e^{\sqrt{2}\tilde{\phi}-4\gamma\psi}(\partial c_1)^2 -\tfrac{1}{2}e^{\sqrt{2}\tilde{\phi} +4\gamma\psi}(\partial c_2)^2 - \Lambda \, \label{Trunc}.
\end{align}
Where the AdS vacuum lives at the following values for the scalars:
\begin{equation}
e^{\phi-\varphi}= |\frac{Q_1}{Q_5}|\,,\qquad \e^{4\bar{\beta\varphi}}= \frac{|Q_1 Q_5|}{4}\,,
\end{equation}
and 
\be
\Lambda=-\frac{32}{|Q_1Q_5|^2}\,.
\ee

What we have described here is a consistent truncation of the bigger action (\ref{generalaction}) that is itself a truncation down to the moduli space of the AdS$_3$ vacuum. The truncation (\ref{Trunc}) is consistent and will be shown below to the generate the solutions of interest by means of $SO(4,5)$ for $\mathbb{T}^4$ or $SO(4,20)$ for $K_3$. 

Interestingly, the two dilaton vectors appearing in the axion kinetic terms of (\ref{Trunc}) are orthogonal since
$2-16\gamma^2 =0$.  This means we have effectively two decoupled $\frac{SL(2,\mathbb{R})}{SO(2)}$ pairs in the truncation. To make this manifest we define
\begin{gather}
\phi_1\equiv \frac{1}{\sqrt{2}} (\tilde{\phi}-\psi) \label{phi1}, \qquad \phi_2\equiv \frac{1}{\sqrt{2}} (\tilde{\phi}+\psi),
\end{gather}
and then (\ref{Trunc}) becomes
\begin{gather}\label{Trunc2}
e^{-1}\mathcal{L}= \mathcal{R}_3-\tfrac{1}{2} (\partial \phi_1)^2 - \tfrac{1}{2} e^{2 \phi_1} (\partial c_1)^2-\tfrac{1}{2} (\partial \phi_2)^2 - \tfrac{1}{2} e^{2 \phi_2} (\partial c_2)^2-\Lambda,
\end{gather}
Note that in Euclidean signature the kinetic terms of $c_1$ and $c_2$ are flipped.

\subsection{No wormholes in the truncation}\label{nowormholes}
For an action in three Euclidean dimensions that consists of decoupled axion-saxion pairs as follows
\begin{gather}
e^{-1}\mathcal{L}= \mathcal{R}_3-\frac{1}{2}\sum_{i=1}^2 \left((\partial \phi_i)^2 - e^{b_i \phi_i} (\partial c_i)^2\right)-\Lambda\,,
\end{gather}
regular wormholes are possible once \cite{ArkaniHamed:2007js}:
\be
\sum_{i=1}^2 \frac{1}{b_i^2} >1\,.\label{AHcond}
\ee
In our case \eqref{Trunc2} we have instead $b_1=b_2=2$ and \emph{do not} satisfy the regularity condition. Hence there are no regular wormholes in this truncation, but neither in the full moduli space since we will later demonstrate that all timelike geodesics have lengths bounded by the ones in the truncation.  This is inconsistent with the claim in \cite{ArkaniHamed:2007js} and we have traced this discrepancy back to a Wick-rotation of ``axion"-fields in \cite{ArkaniHamed:2007js} that were not really independent axion-dilaton pairs in moduli space. \footnote{In a recent paper \cite{Marolf:2021kjc} regular wormholes were found in this setting but they are not the axion wormholes we are considering here. }

\subsection{Uplift to Euclidean D1 strings}
The two axion-dilaton pairs in \eqref{Trunc} exactly source the Euclidean D1 branes wrapping the torus 2-cycles as explained earlier. In here we make this manifest by uplifting the 3D \emph{extremal} instantons.

The 3D metric for the extremal solution is undeformed Euclidean AdS$_3$ and the expressions for the scalars are
\begin{equation}
e^{\phi_i(\rho)}= e^{k_i}(1-a_i\rho), \qquad c_i(\rho)= \pm \frac{e^{-k_i}\,a_i\rho}{1-a_i\rho} +c_{i0}\,,\label{extrsol}
\end{equation}
where $i=1,2$ and $a_i, k_i$ are integration constants. The function $\rho$ is a spherical harmonic function on $AdS_3$ ($\nabla^2\rho=0$), whose details we do not need aside from fixing its shift such that the boundary  of AdS$_3$ (the UV) lives at $\rho=0$, and the IR at $\rho=-\infty$. Regularity requires $a_i>0$. The axion charges are given by:
\be\label{axioncharges}
q_i =\frac{1}{Vol(S^2)}\int_{S^2}\left(e^{2\phi_i}\partial_{\rho}c_i\right)_{\rho=0} = \pm e^{k_i}a_i\,,
\ee
and should be properly quantised.

Now we are ready to uplift the extremal instanton solutions in AdS$_3$. The uplift formula for the metric and the dilaton are\footnote{Here we have fixed $\gamma=1/\sqrt8$. }:
\begin{align}
 2\phi   & =  \phi_1 +\phi_2 + \ln(|Q_1Q_5^{-1}|) \,,\\
2\varphi & =  \phi_1 +\phi_2 -\ln(|Q_1Q_5^{-1}|) \,,\\
\sqrt2 \psi & = \phi_1 -\phi_1\,.
\end{align}
Such that the 10d metric in Einstein becomes:
\begin{equation}
ds^2_{10}  = \left|\frac{Q_1}{Q_5}\right|^{\frac{1}{4}} \left(\left(\frac{h_2}{h_1^3}\right)^{\frac{1}{4}} [\d\theta_1^2+\d\theta_2^2] + \left(\frac{h_1}{h_2^3}\right)^{\frac{1}{4}}[\d\theta_3^2+\d\theta_4^2] \right) + (h_1h_2)^{\frac{1}{4}}  \left|\frac{Q_5}{Q_1}\right|^{\frac{1}{4}} \d s^2_6 \,,
\end{equation}
where the $h_i$ are the following harmonics on AdS$_3$:
\be
h_i = e^{k_i}(1-a_i\rho)\,.
\ee
The dilaton is given by:
\begin{equation}
e^{2\phi}   = \left|\frac{Q_1}{Q_5}\right| h_1h_2\,.
\end{equation}

To compare this with the intersection of Euclidean D1 strings we present the usual supergravity solution for such an intersection based on the harmonic superposition rule and partial smearing \cite{Bergshoeff:1996rn}. In the 10D Einstein frame the solution is given by
\begin{align}
ds^2_{10} = & \left(\frac{H_2}{H_1^3}\right)^{\frac{1}{4}}[\d\tilde{\theta}_1^2+\d\tilde{\theta}_2^2]+ \left(\frac{H_1}{H_2^3}\right)^{\frac{1}{4}}[\d\tilde{\theta}_3^2+\d\tilde{\theta}_4^2] + (H_1H_2)^{\frac{1}{4}}\tilde{\d s^2_6}\,,\\
e^{2\phi} = & e^t H_1H_2\,,
\end{align}
where $H_{1,2}$ are the harmonics of the two Euclidean D1 strings smeared over the $S^3$ and the transversal $\mathbb{T}^2$.
The $e^t$ factor in the dilaton is an integration constant that exists in the case of p-branes in flat non-compact space. Here the background is $AdS_3\times S^3\times T^4$ and that factor is fixed. Its exact value depends on the normalisation of the harmonic functions $H_{1,2}$ which we have not (yet) specified.

The tildes in the above metric are indicating they could be rescaled with respect to the previous normalisation for the metrics on the 4-torus and the 6D space.  Indeed, we find a full match upon identifying $H_i = h_i$, fixing $e^t= |Q_1Q_5^{-1}|$ and rescaling the metrics on the $T^4$ and $\d s_2^6$ by constants involving $Q_1, Q_5$.

\subsection{On-shell actions}
Related, one can demonstrate the the on-shell 3D bulk supergravity action for the instantons equals the on-shell value for the probe Euclidean D1 action in 10D. Let us briefly sketch this.

We first compute the probe action for Euclidean D1 strings in the AdS$_3\times S_3\times \mathbb{T}_4$ vacuum. In 10d string units the DBI action equals
\begin{gather}
S = \frac{n_{1,2}}{g_s} \int_{\Sigma^{1,2}_2} \sqrt{g_2},
\end{gather}
in string frame, with $n_{1,2}$ the number of strings wrapping the two 2-cycles indexed by the labels $1,2$. By moving to Einstein frame we obtain
\begin{gather}\label{probeactions}
S= n_{1,2} e^{-\frac{\phi_0}{2} \pm \frac{\psi_0}{\sqrt{2}}-\frac{\varphi_0}{2}}
\end{gather}
The sign choice for $\psi_0$ determines which of the two 2-cycles inside $T^4$ we wrap the strings around. 
To compute the on-shell action from the backreacted instanton solutions (so beyond probe level) we rely on a well-known fact, reviewed in for instance \cite{Ruggeri:2017grz} that the on-shell action, after holographic renormalisation, is only provided by the total derivative term that one generated from the action in which the axions are dualised to forms. In our set-up, ignoring overall normalisations\footnote{Which would be absorbed in the 3d Planck mass.}, this gives:
\begin{gather}
S^{\text{real}}_{\text{on-shell}} \sim  \int \partial_\rho \left(c_1 e^{2\phi_1} \partial_\rho c_1 + c_2 e^{2\phi_2} \partial_\rho c_2 \right) \sim e^{-k_1}n_1 + e^{-k_2}n_2\,.
\end{gather}
Upon using the uplift formula to rewrite the exponentials in terms of the vevs of the scalars at the boundary\footnote{For example:
$e^{-k_1}=e^{-\phi_1{0}}= e^{-\frac{\phi_0}{2} + \frac{\psi_0}{\sqrt{2}}-\frac{\varphi_0}{2}}$.}, we find a match with the 10D probe actions.

Similarly the imaginary part of the action in 3d should equal the WZ actions in 10d. In 10d it is clear that the instantons have an imaginary part in the action coming from the WZ terms of the Euclidean D1 strings
\be
S_{WZ} =in_{1,2}\int_{\Sigma^{1,2}_2}C_2\,.
\ee
But also in the 3d supergravity the backreacted solutions have an imaginary piece as for instance explained in the appendix of \cite{Bergshoeff:2004pg}. To find the imaginary pieces we need the quantised axion charges \eqref{axioncharges} $q_{1,2}\sim n_{1,2}$ since
\begin{gather}
S^{\text{imaginary}}_{\text{on-shell}} \sim \left(i\, c_1(0) n_1+i \, c_2(0) n_2\right).
\end{gather}
This matches the probe computation on the nose since the axion vevs are, by construction, the $C_2$ form vevs integrated over the internal 2-cycles.

\section{The space of all solutions using group theory}\label{GTSection}
So far we discussed a simple set of solutions corresponding to two stacks of Euclidean D1 branes wrapping the two orthogonal 2-cycles in the internal 4-torus as depicted in \eqref{intersection}. We then showed that these solutions neatly correspond to specific lightlike geodesics on
\be
\left(\frac{SL(2,\mathbb{R})}{SO(1,1)}\right)^2\,,
\ee
which is a consistent truncation of a bigger space corresponding to the proper Wick rotation of
\be\label{modulispace}
\frac{SO(4,m)}{\SO(m)\times\SO(4)}\,,
\ee
with $m=5$ for an internal 4-torus or $m=20$ for a K3 surface. The aim of this section is to clarify what the Wick-rotated moduli space of \eqref{modulispace} is and what the general set of instanton solutions is, by classifying the geodesics on that moduli space, modulo the action of the isometry group ${\rm SO}(4,m)$.
After defining the Wick rotation of the moduli space \eqref{modulispace} we shall characterize the general class of geodesics which the solution described in Sect. \ref{sec:simple} belongs to. This class is characterized by Noether charge matrices which are  nilpotent elements of order three in the defining representation of ${\rm SO}(4,m)$. Solutions of this kind belong to specific \emph{nilpotent orbits} with respect to the action of the isometry group ${\rm SO}(4,m)$. In Sect. \ref{INO} we shall prove that, for generic $m$, \emph{all nilpotent orbits in the coset spaces of the Wick-rotated manifolds always have a representative in the maximally spit universal submanifold defined by $m=4$}.
\par
Therefore, when dealing with extremal solutions, we can restrict our analysis
to the study of light-like geodesics in the Wick-rotated version of \eqref{modulispace} corresponding to the maximal split case $m=4$. 

Let us start with the generic case of
\begin{equation}\label{modulispace2}
\mathscr{ M}=\frac{{\rm SO}(n,n)}{{\rm SO}(n)\times {\rm SO}(n)}\,,
\end{equation}
describing the classical string moduli space of an $n$-torus $T^n$. It is spanned by the internal components $G_{ij},\,B_{ij}$, $i,j=1,\dots, n$, of the metric and of the $B$-field. We can equally think of it as the S-dual moduli space using the $C_2$ field which we use later on. The relation with the AdS moduli space is to be understood as follows. When reducing IIB on the 4-torus we end up with the maximal ungauged six-dimensional supergravity in which the scalar manifold has the form \eqref{modulispace2} with $n=5$.  The moduli $e^{-\frac{\phi}{2}}\,G_{ij},\,C_{ij}$, $i,j=1,\dots,4$ span a submanifold of the form \eqref{modulispace2} with $n=4$, $\phi$ being the ten-dimensional dilaton  and $G_{ij}$ the $\mathbb{T}^4$ metric moduli in the Einstein frame. These 16 moduli are not lifted on the solution of the theory of the form ${\rm AdS}_3\times S^3$, which describe the near horizon geometry of a D1-D5 system. Their moduli space is indeed  a submanifold of the 20-dimensional moduli space of the solution:
\begin{equation}
\frac{{\rm SO}(4,4)}{{\rm SO}(4)\times {\rm SO}(4)} \subset \frac{{\rm SO}(4,5)}{\SO(5)\times\SO(4)}\,,
\end{equation}
where ${\rm SO}(4,5)$ is the stabilizer in ${\rm SO}(5,5)$ of the D1-D5 charge vector \cite{Lu:1997bg, Andrianopoli:2007kz}.\par

\subsection{The Wick-Rotated  $\mathscr{M}^*$}\label{WRM}
Let us denote the Riemannian (i.e. non Wick-rotated) scalar manifold by $\mathscr{M}=G/H$, where $G$ is the isometry group of the form ${\rm SO}(p,q)$ and the isotropy group $H={\rm SO}(p)\times {\rm SO}(q)$ is the maximal compact subgroup of $G$, and the Wick-rotated manifold by $\mathscr{M}^*=G/H^*$, where now $H^*$ is a different (i.e. non-compact) real form of the complexification of $H$.\par 
Let us define the Wick-rotation which is relevant to the problem under consideration. The effect of this rotation is to change the sign of the metric on the manifold along the directions of the axion fields. These scalars can be characterized as parameters of a maximal abelian subalgebra $\mathfrak{A}$, consisting of nilpotent generators, of the isometry one $\mathfrak{g}=\mathfrak{so}(p,q)$ \cite{Andrianopoli:1996zg,Cremmer:1997ct}. Let $\theta$ denote the Cartan involution on $\mathfrak{g}$, which defines its Cartan decomposition:
\begin{equation}
    \mathfrak{g}=\mathfrak{K}\oplus \mathfrak{H}\,,
    \label{cartandec}
\end{equation}
into the space of the non-compact generators (i.e. hermitian in a suitable basis) $\mathfrak{K}$ and the maximal compact subalgebra $\mathfrak{H}=\mathfrak{so}(p)\oplus \mathfrak{so}(q)$ ($\theta(\mathfrak{K})=-\mathfrak{K},\,\theta(\mathfrak{H})=\mathfrak{H}$). The space $\mathfrak{K}_{2}\equiv \mathfrak{A}-\theta(\mathfrak{A})$ is a subspace of $\mathfrak{K}$ while $\mathfrak{H}_{2}\equiv \mathfrak{A}+\theta(\mathfrak{A})$ is contained in $\mathfrak{H}$. The grading properties defining $\mathfrak{A}$ imply that the spaces $\mathfrak{K}\,, \mathfrak{H}$ decompose as follows:
\begin{equation}
    \mathfrak{K}=\mathfrak{K}_1\oplus \mathfrak{K}_2\,\,,\,\,\,\, \mathfrak{H}=\mathfrak{H}_1\oplus \mathfrak{H}_2\,,\label{HK}
\end{equation}
where $\mathfrak{H}_1$ is a subalgebra of $\mathfrak{H}$ generating a subgroup $H_c\subset H$, $H_c=e^{\mathfrak{H}_1}$. Under the adjoint action of $H_c$, the space $\mathfrak{H}_2$ transforms in a representation $\mathcal{R}$. The Wick rotation is effected by interchanging in (\ref{HK}) the spaces $\mathfrak{K}_2$ and $\mathfrak{H}_2$ so as to define:
\begin{equation}
      \mathfrak{K}^*=\mathfrak{K}_1\oplus \mathfrak{H}_2\,\,,\,\,\,\, \mathfrak{H}^*=\mathfrak{H}_1\oplus \mathfrak{K}_2\,,\label{HK*}
\end{equation}
where now $\mathfrak{K}^*$ is the coset space of the Wick-rotated manifold $\mathscr{M}^*$, isomorphic to its tangent space at the origin, while the algebra $\mathfrak{H}^*$ generates its non-compact isotropy group $H^*$. The decomposition of $\mathfrak{g}$ into $\mathfrak{K}^*,\,\mathfrak{H}^*$:
\begin{equation}
    \mathfrak{g}=\mathfrak{H}^*\oplus \mathfrak{K}^*\,,\label{pcartan}
\end{equation}
is referred to as \emph{pseudo-Cartan} decomposition. These two spaces are now eigenspaces of a new involution $\theta^*$: $\theta^*(\mathfrak{H}^*)=\mathfrak{H}^*,\,\theta^*(\mathfrak{K}^*)=-\mathfrak{K}^*$.
The metric on the tangent space at the origin of $\mathscr{M}^*$ is defined by the restriction of the Cartan-Killing metric of $\mathfrak{g}$ to $ \mathfrak{K}^*$ and thus has negative signature directions along a basis of $\mathfrak{H}_2$. These are the directions of the axionic fields since only the axionic isometry generators have components in $\mathfrak{H}_2$. In particular the axion charges are defined as the components of the Noether charge matrix $Q$ of a geodesic along the generators of $\mathfrak{H}_2$.\par
As far as the $\mathfrak{g}=\mathfrak{so}(p,q)$ algebra is concerned, there are two kinds of maximal abelian subalgebras which are relevant to our discussion.
\begin{itemize}
\item[i)]{A generic $\mathfrak{so}(p,q)$ algebra always has a $(p+q-2)$-dimensional maximal abelian subalgebra defined by the decomposition:
\begin{equation}
 \mathfrak{so}(p,q)=\mathfrak{so}(1,1)_0\oplus\mathfrak{so}(p-1,q-1)_0  \oplus ({\bf p+q-2})_{+1}\oplus \overline{({\bf p+q-2})}_{-1}\,,
\end{equation}
where the grading refers to the $\mathfrak{so}(1,1)$-generator. Since there are no generators with grading $+2$ or $-2$, the subspaces in the representations $({\bf p+q-2})_{+1}$ and $ \overline{({\bf p+q-2})}_{-1}$ are separately abelian subalgebras. In this case we can choose $\mathfrak{A}=({\bf p+q-2})_{+1}$. An example of this subalgebra is the one parametrized by the eight R-R scalars $C_{ij},\,C_{ijkl},\,C_{(0)}$ within $\mathfrak{so}(5,5)$ in the maximal $D=6$ theory originating from Type IIB superstring compactified on $T^4$. Another instance of such abelian subalgebra is the one parametrized by the 22 components $C_I$ of the Type IIB R-R 2-form $C_{(2)}$ along the 2-cycles of an internal $K_3$. In this case the isometry group of the moduli space is ${\rm SO}(4,20)$.}
\item[ii)]{Only for $p=q=n$ we have a maximal abelian subalgebra of dimension $n(n-1)/2$ defined by the following decomposition:
\begin{equation}
    \mathfrak{so}(n,n)=\mathfrak{so}(1,1)_0\oplus\mathfrak{sl}(n)_0  \oplus \left({\bf \frac{n(n-1)}{2}}\right)_{+1}\oplus \overline{\left({\bf \frac{n(n-1)}{2}}\right)}_{-1}\,,
\end{equation}
The same grading argument used in case $i)$ implies that the subspaces of generators with gradings $+1$ and $-1$ are separately abelian subalgebras.
In this case $\mathfrak{A}=\left({\bf \frac{n(n-1)}{2}}\right)_{+1}$ and an explicit construction of its generators, as $2n\times 2n$ matrices in a suitable basis, is given below in eq. (\ref{Aii}). Instances of this subalgebra is the one parametrized by the moduli $B_{ij}$ in the algebra $\mathfrak{so}(n,n)$ acting on the moduli $G_{ij},\,B_{ij}$ of Type IIB supergravity compactified on $T^n$, or by the moduli $C_{ij}$ within the $\mathfrak{so}(n,n)$ acting, in the same $D=6$ theory, on the moduli $G_{ij},\,C_{ij}$. When $n=4$ this maximal abelian subalgebra is isomorphic to the one in case $i)$, having both dimension $6$. They are related by triality.}
\end{itemize}
In case $i)$ the Wick-rotated manifold is:
\begin{equation}
    \mathscr{M}^*=\frac{{\rm SO}(p,q)}{{\rm SO}(1,p-1)\times {\rm SO}(1,q-1)}\,,
\end{equation}
the group $H_c$ is ${\rm SO}(p-1)\times {\rm SO}(q-1)$ and the representation $\mathcal{R}$ in which $\mathfrak{K}_2,\,\mathfrak{H}_2$ transform under the adjoint action of $H_c$ is the ${\bf (p-1,1)}\oplus {\bf (1,q-1)}$. We can therefore view $\mathfrak{K}_2$ as the coset space of the following symmetric manifold:
\begin{equation}
    \frac{{\rm SO}(1,p-1)}{{\rm SO}(p-1)}\times \frac{{\rm SO}(1,q-1)}{{\rm SO}(q-1)}=e^{\mathfrak{K}_2}\,.
\end{equation}\par
In case $ii)$ the Wick-rotated manifold is:
\begin{equation}
    \mathscr{M}^*=\frac{{\rm SO}(n,n)}{{\rm SO}(n,\mathbb{C})}\,,\label{modulispace3}
\end{equation}
 $H_c={\rm SO}(n)$ and $\mathcal{R}={\bf \frac{n(n-1)}{2}}$.\par
\par
Below we shall expand on the case $ii)$ and study the geometry of the Wick-rotated manifold. The manifold is parametrized by the moduli $\tilde{G}_{ij}=e^{-\phi/2}\,G_{ij},\,C_{ij}$ and the Wick-rotation flips the sign of the kinetic terms of $C_{ij}$.  We are interested in the $n=4$ case, for which  ${\rm SO}(4,\mathbb{C})\sim {\rm SL}(2,\mathbb{C})^2\sim {\rm SO}(1,3)^2$. \par
Let us use, as ${\rm SO}(n,n)$-invariant metric in the defining representation, the matrix:
\begin{equation}
\eta=\left(\begin{matrix}{\bf 0} & {\bf 1}\cr {\bf 1} & {\bf 0}\end{matrix}\right)=\sigma_1\otimes {\bf 1}_n\,,
\end{equation}
where ${\bf 1}_n$ is the $n\times n$ identity matrix and $\sigma_1,\,\sigma_2,\,\sigma_3$ are the Pauli matrices. According to the Cartan decomposition \eqref{cartandec}, the isometry algebra $\mathfrak{g}=\mathfrak{so}(n,n)$ splits into its maximal compact subalgebra
where $\mathfrak{H}=\mathfrak{so}(n)\oplus \mathfrak{so}(n)$ and the space $\mathfrak{K}$ consisting of the hermitian matrices in the algebra $\mathfrak{g}$. According to our discussion above, 
we can further split the subspaces $\mathfrak{H}$ and $\mathfrak{K}$ as in \eqref{HK}, where $\mathfrak{K}_2,\,\mathfrak{H}_2$ are spanned, respectively, by the hermitian and anti-hermitian components of the elements of the maximal abelian subalgebra $\mathfrak{A}$. In the ${\rm SO}(n,n)$ defining representation the generic representatives of the above subspaces have the following form:
\begin{align}
\mathfrak{H}_1&=\{{\bf 1}_2\otimes {\bf A}\}\,\,,\,\,\,\mathfrak{H}_2=\{\sigma_1\otimes {\bf A}'\}\,,\nonumber\\
\mathfrak{K}_1&=\{\sigma_3\otimes \boldsymbol{\gamma}\}\,\,,\,\,\,\mathfrak{K}_2=\{i\,\sigma_2\otimes {\bf C}\}\,,
\end{align}
the matrices ${\bf A},\,{\bf A}',\,{\bf C}$ being generic $n\times n$ antisymmetric matrices and $\boldsymbol{\gamma}$ being  a generic symmetric matrix. The subspace $\mathfrak{K}_1$ is the coset-space of the metric moduli $G_{ij}$ of $T^n$, suitably combined with the ten-dimensional dilaton $\phi$, and it generates the submanifold $\frac{{\rm GL}(n,\mathbb{R})}{{\rm SO}(n)}$, and is spanned by $\boldsymbol{\gamma}=(\gamma_{ij})=\boldsymbol{\gamma}^T$.
As discussed above, the Wick rotation is effected by exchanging the roles of the spaces $\mathfrak{H}_2$ and $\mathfrak{K}_2$, so that the algebra $\mathfrak{g}$
decomposes according to the  pseudo-Cartan decomposition (\ref{pcartan}), 
where $\mathfrak{H}^*,\,\mathfrak{K}^*$ are given in \eqref{HK*}.
Now $\mathfrak{H}^*$ is the algebra $\mathfrak{so}(n,\mathbb{C})$ while $\mathfrak{K}^*$ has $n(n-1)/2$ negative signature directions corresponding to the compact generators in $\mathfrak{H}_2$.
The pseudo-Cartan decomposition is defined by an involution $\theta^*$, defined by the matrix $\eta'=\sigma_3\otimes {\bf 1}_n$ as follows:
\begin{equation}
\theta^*(\mathfrak{H}^*)=-\eta'\,(\mathfrak{H}^*)^T\,\eta'=\mathfrak{H}^*\,\,,\,\,\,\,\theta^*(\mathfrak{K}^*)=-\eta'\,(\mathfrak{K}^*)^T\,\eta'=
-\mathfrak{K}^*\,.
\end{equation}
\par
We then have the following local isometric representation:
\begin{equation}
\frac{{\rm SO}(n,n)}{{\rm SO}(n,\mathbb{C})}\sim \left(\frac{{\rm GL}(n,\mathbb{R})}{{\rm SO}(n)}\right)\ltimes e^{\mathfrak{A}}\,,\label{isomet}
\end{equation}
where $\mathfrak{A}$ is the abelian algebra generated by nilpotent matrices parametrized by ${\bf C}=(C_{ij})$ while $\frac{{\rm GL}(n,\mathbb{R})}{{\rm SO}(n)}$ is spanned by $\boldsymbol{\gamma}$, related to the metric moduli of the internal torus. We can use the following matrix representations:
\begin{equation}
\mathfrak{A}=\{\sigma_+\otimes {\bf C}=\sigma_+\otimes \frac{1}{2}\,t^{ij}\,C_{ij}\}\,,\label{Aii}
\end{equation}
where $\sigma_+\equiv (\sigma_1+i\,\sigma_2)/2$ satisfies the relation $[\sigma_3,\,\sigma_+]=2 \sigma_+$, while $(t^{ij})_{kl}=2\,\delta^{ij}_{kl}$.
According to (\ref{isomet}) we define the coset representative as follows:
\begin{equation}
L=e^{\mathfrak{A}}\,L_G\,,
\end{equation}
where $L_G\in e^{\mathfrak{K}_1}$ is the coset representative of $\frac{{\rm GL}(n,\mathbb{R})}{{\rm SO}(n)}$. The matrix $\mathcal{M}$ locally describing the coset is defined as follows:
\begin{equation}
\mathcal{M}\equiv L\,\eta'\,L^T=e^{\mathfrak{A}}\,L_G\,\eta'\,L_G^T\,\left(e^{\mathfrak{A}}\right)^T=e^{\mathfrak{A}}\,\mathcal{M}_G\eta'\,\left(e^{\mathfrak{A}}\right)^T\,,
\end{equation}
where $\mathcal{M}_G\equiv L_G\,L_G^T$ and we have used the property that $L_G$ commutes with $\eta'$.\par
The generic element of the group $e^{\mathfrak{A}} $ and $\mathcal{M}_G$ have the form:
\begin{equation}
e^{\mathfrak{A}}=\left(\begin{matrix}{\bf 1} & {\bf C}\cr {\bf 0} & {\bf 1}\end{matrix}\right)\,,\,\,\,\,\mathcal{M}_G=\left(\begin{matrix}\tilde{{\bf G}} & {\bf 0}\cr {\bf 0} & \tilde{{\bf G}}^{-1}\end{matrix}\right)\,,
\end{equation}
where $\tilde{{\bf G}}=(\tilde{G}_{ij})\equiv e^{2\,\boldsymbol{\gamma}} $.\par
The matrix $\mathcal{M}$ reads:
\begin{equation}
\mathcal{M}=\left(\begin{matrix}{\bf 1} & {\bf C}\cr {\bf 0} & {\bf 1}\end{matrix}\right)\left(\begin{matrix}{\bf 1} & {\bf 0}\cr {\bf 0} & -{\bf 1}\end{matrix}\right)\left(\begin{matrix}\tilde{{\bf G}} & {\bf 0}\cr {\bf 0} & \tilde{{\bf G}}^{-1}\end{matrix}\right)\left(\begin{matrix}{\bf 1} & {\bf 0}\cr -{\bf C} & {\bf 1}\end{matrix}\right)=\left(\begin{matrix}\tilde{{\bf G}}+ {\bf C}\,\tilde{{\bf G}}^{-1}\, {\bf C}& -{\bf C}\tilde{{\bf G}}^{-1}\cr \tilde{{\bf G}}^{-1}\,{\bf C} & -\tilde{{\bf G}}^{-1}\end{matrix}\right)\,.\label{formM}
\end{equation}
From this we can compute the metric on moduli space as:
\begin{equation}
ds^2=\frac{1}{4}\,{\rm Tr}\left[\mathcal{M}^{-1}\,d\mathcal{M}\,\mathcal{M}^{-1}\,d\mathcal{M}\right]=\frac{1}{2}\,\left(\tilde{G}^{mp}\tilde{G}^{nq}d\tilde{G}_{mn} d\tilde{G}_{pq}-\tilde{G}^{mp}\tilde{G}^{nq}dC_{mn} dC_{pq}\right)\,,
\end{equation}
where $$\tilde{G}_{ij}=e^{-\frac{\phi}{2}}\,G_{ij}\,,$$
$G_{ij}$ being the metric of the 4-torus in the Einstein frame. \footnote{The combination $e^\phi\,{\rm det}(G_{ij})^{\frac{1}{2}}$ is fixed in terms of the D1-D5 charges.}
The sigma-model Lagrangian density then reads:
\begin{equation}
\mathcal{L}_{(\tilde{G},C)}=-\frac{1}{4}\left(\tilde{G}^{mp}\tilde{G}^{nq}\partial_\mu \tilde{G}_{mn} \partial^\mu \tilde{G}_{pq}-\tilde{G}^{mp}\tilde{G}^{nq}\partial_\mu C_{mn} \partial^\mu C_{pq}\right)\,.
\end{equation}
We see that indeed the axion scalars have the opposite sign of the kinetic term.  In what follows we use the exponential map to solve and classify the geodesics equations. In practice this means that the above sigma model can trivially be solved for geodesics in terms of the symmetric coset matrix $\mathcal{M}$. Let us, for the sake of notational simplicity, collectively denote the moduli $\tilde{G}_{ij},\,C_{ij}$ by $\phi^I$.
 The geodesics on $\mathscr{M}^*$ can be classified in orbits with respect to the action of the isometry group $G$. More precisely, using transformations in $G/H^*$, the initial point at $\rho=0$ can always be chosen to coincide with a given one $\phi_0=(\phi_0^I)$. Once the this point is fixed
 we still have the freedom of changing the \emph{initial velocity}, represented by the Noether charge matrix $Q_0$, within the tangent space to the moduli space at $\phi_0$, by means of the isotropy group  $H_{\phi_0}^*$ of $\phi_0$. For the sake of simplicity we can start fixing the initial point to be the origin $$\phi_0=O\,\,\Leftrightarrow\,\,\,\,\tilde{G}_{ij}(\rho=0)=\delta_{ij},\,C_{ij}(\rho=0)=0\,,$$ so that $H_{\phi_0}^*=H^*$ and  the geodesics are completely determined by the ``initial velocity''  $Q$, now element of $\mathfrak{K}^*$. The geodesic is solution to the matrix equation:
\begin{equation}
\mathcal{M}\left(\phi(\rho)\right)=\mathcal{M}\left(\tilde{{\bf G}}(\rho),\,{\bf C}(\rho)\right)=\eta'\cdot e^{2Q\rho}\,.\label{matrixGeq}
\end{equation}
As an element of $\mathfrak{K}^*$, the general form of $Q$ is:
\begin{equation}
Q=\sigma_3\otimes \boldsymbol{\gamma}+\sigma_1\otimes {\bf c}\,,\label{genQ}
\end{equation}
where $\boldsymbol{\gamma}^t=\boldsymbol{\gamma}$ and ${\bf c}^t=-{\bf c}$.\par
The geodesic $\phi(\rho,\phi_0)$ through a generic point $\phi_0$ at $\rho=0$ is then obtained from the one through the origin  by solving the matrix equation:
\begin{equation}
   \mathcal{M}(\phi(\rho,\phi_0))=L(\phi_0)\, \mathcal{M}(\phi(\rho))\, L(\phi_0)^T=\mathcal{M}(\phi_0)\,e^{2\rho\,Q_0}\,,
\end{equation}
where $Q_0\equiv L(\phi_0)^{-1\,T}\,Q\,L(\phi_0)^{T}$ is an element of the tangent space to the moduli space at $\phi_0$.

\subsection{The general solution for the geodesics}\label{totalsolution}

Let us now describe the general form of the geodesics in $\mathscr{M}^*$ generated by a Noether charge matrix $Q\in \mathfrak{K}^*$, through the origin. They belong to the three classes:
\begin{enumerate}
    \item Extremal instantons. These are the lightlike geodesics and then the $Q$-matrix is necessarily nilpotent. As we shall prove in Section \ref{INO}, the maximal degree of nilpotency of a nilpotent element $Q$ of $\mathfrak{K}^*$, in the representation ${\bf 8}_v$ of ${\rm so}(4,4)$, is four: $Q^4={\bf 0}$. The extremal solutions constructed in Section 3 are generated by an order-2 nilpotent matrix $Q$;
    \item Over-extremal instantons. These are the timelike geodesics and correspond to wormholes, but they will not be regular in their scalar profiles as we explained before. Then $Q$ is semisimple with imaginary eigenvalues. As we are interested in evaluating the maximal length of timelike geodesics, we can take $Q$ in $\mathfrak{H}_2$.
    \item Sub-extremal instantons. These are the spacelike geodesics with $Q$ having  real eigenvalues in $\mathfrak{K}_1$.
\end{enumerate}
The regularity condition on the above solutions is:
\begin{equation}
    \infty>\tilde{G}_{ij}>0\,.\label{regular}
\end{equation}
\paragraph{Extremal solutions.} Since, in the representation we are currently considering, the Noether charge matrix $Q$ is nilpotent of order at most 4, we give the explicit form of the generic solution in the $Q^4= {\bf 0}$ orbit. 
The two matrices $\boldsymbol{\gamma},\,{\bf c}$ satisfy the conditions:
\begin{equation}
    (\boldsymbol{\gamma}^2+{\bf c}^2)^2-[\boldsymbol{\gamma},\,{\bf c}]^2={\bf 0}\,\,,\,\,\, (\boldsymbol{\gamma}^2+{\bf c}^2)\cdot [\boldsymbol{\gamma},\,{\bf c}]=-[\boldsymbol{\gamma},\,{\bf c}]\cdot (\boldsymbol{\gamma}^2+{\bf c}^2)\,.
\end{equation}
The general form of the geodesic is:
\begin{align}
\tilde{{\bf G}}(\rho)&=(\tilde{G}_{ij}(\rho))=\left({\bf 1}-2 \rho \,\boldsymbol{\gamma}+2\rho^2\, (\boldsymbol{\gamma}^2+{\bf c}^2)-\frac{4}{3}\,\left((\boldsymbol{\gamma}^2+{\bf c}^2)\cdot \boldsymbol{\gamma} +[\boldsymbol{\gamma},\,\boldsymbol{c}]\cdot \boldsymbol{c}\right)\rho^3\right)^{-1}\,,\nonumber\\
{\bf C}(\rho)&=-2 \rho\,\tilde{G}(\rho)\cdot \left({\bf c}- \rho \,[\boldsymbol{\gamma},\,{\bf c}]+\frac{2}{3}\rho^2 \,\left((\boldsymbol{\gamma}^2+{\bf c}^2)\cdot \boldsymbol{c} -[\boldsymbol{\gamma},\,\boldsymbol{c}]\cdot \boldsymbol{\gamma}\right)\right)\,.\label{solextregen}
\end{align}
The matrices $\boldsymbol{\gamma}$ and $\boldsymbol{c}$ are constrained by the regularity condition (\ref{regular}).\par
If $Q$ belongs to the $Q^3$-orbit, the following conditions hold:
\begin{equation}
(\boldsymbol{\gamma}^2+{\bf c}^2)\cdot \boldsymbol{\gamma}+[\boldsymbol{\gamma},\,{\bf c}]\cdot {\bf c}={\bf 0}\,,\,\,\,(\boldsymbol{\gamma}^2+{\bf c}^2)\cdot {\bf c}-[\boldsymbol{\gamma},\,{\bf c}]\cdot \boldsymbol{\gamma}={\bf 0}\,,
\end{equation}
which set the $\rho^3$ terms in the solution (\ref{solextregen}) to zero.
Finally, if $Q^2=0$ we have the stronger condition:
\begin{equation}
\boldsymbol{\gamma}^2+{\bf c}^2={\bf 0}\,,\,\,\,[\boldsymbol{\gamma},\,{\bf c}]={\bf 0}\,,
\end{equation}
and also the $\rho^2$ terms in (\ref{solextregen})  vanish. We shall discuss a normal form for a $Q$ in this orbit in subsection \ref{normal}.
\paragraph{Semisimple $Q$ in $\mathfrak{K}_1$.}
Consider now $Q$ semisimple in $\mathfrak{K}_1$. It has real eigenvalues. This is the case if we set ${\bf c}={\bf 0}$ so that $Q=\sigma_3\otimes \boldsymbol{\gamma}$ in the coset space of ${\rm GL}(4,\mathbb{R})/{\rm SO}(4)$. The general geodesic has the form:
\begin{equation}
\tilde{{\bf G}}(\rho)=\cosh(2\rho \, \boldsymbol{\gamma})+\sinh(2\rho \, \boldsymbol{\gamma})\,\,,\,\,\,{\bf C}(\rho)={\bf 0}\,,
\end{equation}
The inverse of $\tilde{G}$ is $\tilde{G}^{-1}(\rho)=\cosh(2\rho \, \boldsymbol{\gamma})-\sinh(2\rho \, \boldsymbol{\gamma})$.
This matrix can be diagonalized by an ${\rm SO}(4)$ rotation. If we denote by $\gamma_i$ the eigenvalues of $\boldsymbol{\gamma}$, in the basis in which this matrix is diagonal, so is the metric and reads:
\begin{equation}
\tilde{G}_{ij}(\rho)=\delta_{ij}\,(\cosh(2\rho \, \gamma_i)+\sinh(2\rho \, \gamma_i))\,.
\end{equation}
\paragraph{Semisimple $Q$ in $\mathfrak{H}_2$.}
Consider now $Q$ semisimple in $\mathfrak{H}_2$. It has imaginary eigenvalues. This is the case if we set $\boldsymbol{\gamma}={\bf 0}$ so that $Q=\sigma_1\otimes \boldsymbol{c}$. The general geodesic has the form:
\begin{equation}
\tilde{{\bf G}}(\rho)=\cosh(2\rho \, {\bf c})^{-1}\,\,,\,\,\,{\bf C}(\rho)=\sinh(2\rho \, {\bf c})\cdot\cosh(2\rho \, {\bf c})^{-1} \,.
\end{equation}
By means of an ${\rm SO}(4)$ rotation, ${\bf c}$ can be brought to a skew-diagonal form ${\bf c}_{SD}$, with only non vanishing entries $c_{12}$ and $c_{34}$:
\begin{equation}
{\bf c}_{SD}=\left(\begin{matrix}0& c_1  & 0 & 0\cr -c_1 & 0 & 0& 0\cr 0&0&0&c_2\cr 0&0&-c_2&0\end{matrix}\right)\,.
\end{equation}
In this basis the solution is characterized by the following only non-vanishing components of $\tilde{G}(\rho)$ and $C(\rho)$:
\begin{align}
\tilde{G}_{11}(\rho)&=\tilde{G}_{22}(\rho)=\cos(2\rho c_1)^{-1}\,\,,\,\,\,\tilde{G}_{33}(\rho)=\tilde{G}_{44}(\rho)=\cos(2\rho c_2)^{-1}\,,\nonumber\\
C_{12}(\rho)&=-C_{21}(\rho)=\tan(2\rho c_1)\,\,,\,\,\,C_{34}(\rho)=-C_{43}(\rho)=\tan(2\rho c_2)\,.
\end{align}
This solution generates the most general timelike geodesic. It belongs to the truncation considered in Section \ref{sec:simple} and describes singular wormholes.

\par
So far we have been working with the ${\bf 8}_v$ representation of ${\rm SO}(4,4)$ which branches with respect to ${\rm GL}(4,\mathbb{R})$ as ${\bf 8}_v\rightarrow {\bf 4}_++\bar{{\bf 4}}_-$. When embedding the defining representation of ${\rm SO}(4,4)$ within ${\rm SO}(4,m) $, $m>4$, we shall be working with the representation ${\bf 8}_s$ instead, related to ${\bf 8}_v$ by triality, which branches with respect to the same subgroup as ${\bf 8}_s\rightarrow {\bf 6}_0+{\bf 1}_-+{\bf 1}_+$. The maximal abelian subalgebra $\mathfrak{A}$ will then be of kind $i)$ instead of $ii)$ and some of the allowed nilpotent orbits for $Q$ will change accordingly.
\subsection{The Issue of Nilpotent Orbits}\label{INO}
Extremal solutions are described by a nilpotent Noether charge matrix $Q$ in $\mathfrak{K}^*$ which is then classified in orbits with respect to the adjoint action of $H^*$. In the previous sections we focused on the geodesic solutions in a moduli space of the form (\ref{modulispace}) with $m=4$. Here we discuss how general this choice is and prove that the nilpotent orbits of $Q$ in the moduli space with $m=5$ all have a representative in the maximally split subspace with $m=4$. We shall refrain from reviewing the theory of nilpotent orbits of a semisimple Lie group, for which we refer the reader to 
some useful reviews \cite{collingwood,Dietrich:2013non}. The nilpotent orbits in $\mathfrak{so}(p,q)$ were classified in \cite{Djokovic}.
The general problem which is relevant to our analysis is that of studying the nilpotent orbits within $\mathfrak{K}^*$ with respect to the adjoint action of $H^*$. This problem is referred to, in the mathematical literature, as that of classifying the nilpotent orbits of the vector space $\mathfrak{K}^*$ 
associated with the real semisimple symmetric pair $(\mathfrak{g},\,\mathfrak{H}^*)$. For the sake of concreteness we shall consider $\mathfrak{g}=\mathfrak{so}(p,q)$. In the case $p=q=4$ and $\mathfrak{H}^*=\mathfrak{sl(2,\mathbb{R})^4}$ the problem was solved in \cite{Dietrich:2016ojx,Ruggeri:2016dfk}. However the real semisimple symmetric pair which is relevant to our present analysis is the one with $\mathfrak{g}=\mathfrak{so}(p,q)$ and $\mathfrak{H}^*=\mathfrak{so}(1,p-1)\oplus \mathfrak{so}(1,q-1)$, for the special values $p=4,\,q=m$. Here we shall limit ourselves to identifying, in the latter case, those $G$-nilpotent orbits which have a representative in $\mathfrak{K}^*$, without further splitting them with respect to the action of $H^*$.\par
 According to the Jacobson-Morozov theorem \cite{collingwood}, any nilpotent element $e$ of a real Lie algebra $\mathfrak{g}$ can be thought of as
part of a \emph{standard triple} of $\mathfrak{sl}(2,\mathbb{R})$-generators $\{h,e,f\}$ satisfying the standard commutation relations
$$[h,e]=e\,\,,\,\,\,[h,f]=-f\,\,,\,\,\,\,[e,f]=h\,.$$
We are interested in nilpotent elements $e$ which lie in the coset space $\mathfrak{K}^*$. Then the standard triple can be chosen so that $f\in \mathfrak{K}^*$ and $h$ be a non-compact generator in $\mathfrak{H}^*$. It is known that the nilpotent orbits in the complexification $\mathfrak{g}^\mathbb{C}$ of $\mathfrak{g}$ with respect to $G^\mathbb{C}=\exp{\mathfrak{g}^\mathbb{C}}$ are defined by the inequivalent embeddings of the $\mathfrak{sl}(2,\mathbb{C})={\rm Span}(h,e,f)$ inside $\mathfrak{g}^\mathbb{C}$, which in turn are defined by the different decompositions of the defining representation of 
$G^\mathbb{C}$ with respect to the corresponding ${\rm SL}(2,\mathbb{C})$ group  (with a certain multiplicity prescription). Each of these decompositions is characterized by a partition of the dimension of the ${\bf (p+q)}$ representation of $\mathfrak{g}^\mathbb{C}=\mathfrak{so}(p,q;\mathbb{C})$. If, with respect to ${\rm SL}(2,\mathbb{C})$, the ${\bf (p+q)}$ representation branches as follows:
\begin{equation}
    {\bf (p+q)}\rightarrow \bigoplus_{i=1}^\ell {\rm k}_i\times[{\bf s}_i]\,,
\end{equation}
where we have used the ordering $s_\ell\ge s_{\ell-1}\ge \dots \ge s_1$,
the partition is denoted by $ [(2s_\ell+1)^{{\rm k}_\ell},\dots,(2s_1+1)^{{\rm k}_1}] $ and represented by a corresponding Young tableau. According to the general theory only certain partitions can occur and with certain multiplicities. When we consider real nilpotent orbits there is a finer structure and each nilpotent ${\rm SO}(p,q)$-orbit in $\mathfrak{so}(p,q)$ is described by a \emph{graded Young tableau} \cite{collingwood}. The order of nilpotency of the corresponding orbit in the defining representation is $2s_\ell+1$ since the $h$-grading of the element $e$ of the orbit is $1$ and the minimal and maximal eigenvalues of $h$ in the defining representation are $-s_\ell$ and $s_\ell$, respectively. For $\mathfrak{so}(4,4;\mathbb{C})$ the partitions are:
\begin{equation}[1^8],\,[2^2,1^4],\,[3,1^5],\,[2^4]^I,\,[2^4]^{II},\,[3,2^2,1],\,[3^2,1^2],\,[5,1^3],\,[4^2]^I,\,[4^2]^{II},\,[5,3],\,[7,1]\,,\label{urbietorbi}\end{equation}
$[1^8]$ being the trivial orbit corresponding to the zero-matrix. The orbits $[3,1^5],\,[2^4]^I,\,[2^4]^{II}$ are related to one another by ${\rm SO}(4,4)$-triality and so are the orbits $[5,1^3],\,[4^2]^I,\,[4^2]^{II}$.
We choose the embedding ${\rm SO}(4,4)$ inside ${\rm SO}(4,m)$ to be such that the defining representation ${\bf 4+m}$ of the latter, when branched with respect to the former, contains the ${\bf 8}_s$ representation instead of the  ${\bf 8}_v$. The difference is that, with respect to the ${\rm GL}(4,\mathbb{R})$ group acting on the metric moduli of the 4-torus, the two 8-dimensional representations branch differently: ${\bf 8}_s\rightarrow {\bf 6}_0+{\bf 1}_-+{\bf 1}_+,\,{\bf 8}_v\rightarrow {\bf 4}_++\overline{{\bf 4}}_-$. This choice of the embedding of ${\rm SO}(4,4)$ inside ${\rm SO}(4,m)$ is  appropriate to the problem at hand 
since if we consider the chain of embeddings ${\rm SO}(4,4)\subset {\rm SO}(4,5)\subset {\rm SO}(5,5)$, ${\rm SO}(5,5)$ being the global symmetry group of the maximal six-dimensional supergravity, when branching the ${\bf 10}$ of the latter, describing the 3-form field strengths, with respect to ${\rm SO}(4,4)\times {\rm SO}(1,1)$ we have ${\bf 10}\rightarrow {\bf 8}_{s\,0}+{\bf 1}_{+}+{\bf 1}_{-}$, since the ${\bf 8}_{s\,0}$ contains the six 3-forms $H_{ij\,\mu\nu\rho}$ in the ${\bf 6}_0$ of ${\rm GL}(4,\mathbb{R})$. For the same reason the branching of the adjoint representation of ${\rm SO}(5,5)$ with respect to ${\rm SO}(4,4)$ contains the ${\bf 8}_s$ instead of the ${\bf 8}_v$. In the previous sections we have being working with the ${\rm SO}(4,4)$-generators in the ${\bf 8}_v$. Now we shall use the ${\bf 8}_s$ representation of the same group instead. This will affect the orbit assignment of a nilpotent generator in $\mathfrak{so}(4,4)$: a generator in the orbits $[2^4]^I,\,[2^4]^{II}$ as a matrix in the representations ${\bf 8}_v$ or ${\bf 8}_c$, in the ${\bf 8}_s$ will belong to the orbit $[3,1^5]$. Similarly triality will map the orbits $[4^2]^I$ or $[4^2]^{II}$, when the nilpotent generator is in the ${\bf 8}_v$ or ${\bf 8}_c$, into the orbit $[5,1^3]$ when it is represented in the ${\bf 8}_s$.\par
The main observation is that if the neutral element $h$ of the standard triple associated with a nilpotent generator $e\in \mathfrak{K}^*$ can always be chosen to lie in the subspace $\mathfrak{K}_2\in \mathfrak{H}^*$. It then  transforms under the adjoint action of $H_c={\rm SO}(p-1)\times {\rm SO}(q-1)\subset H^*$ in the representation $({\bf p-1,1})\oplus ({\bf 1,q-1})$. Restricting to $p=4$ and $q=m$, by acting on the whole triple by means of the compact symmetry group $H_c$, $h$ can always be rotated into a minimal two-dimensional subspace $\mathfrak{K}_2^{(N)}={\rm Span}(\mathcal{J}_\ell)_{\ell=1,2}$ of $\mathfrak{K}_2$ which is contained in the subalgebra $\mathfrak{so}(4,4)$ of $\mathfrak{so}(4,m)$. This subspace defines the non-compact rank of the coset $\frac{{\rm SO}(1,3)}{{\rm SO}(3)}\times \frac{{\rm SO}(1,3)}{{\rm SO}(3)}$. The reason behind this is that any $n$-vector ${\bf v}$ in the defining representation of ${\rm SO}(n)$ can be rotated by means of this group in the normal form: ${\bf v}=(\pm |{\bf v}|,0,\dots,0)$. Thus we can always rotate a generic $h$ in the coset space $\mathfrak{K}_2$ of $\frac{{\rm SO}(1,3)}{{\rm SO}(3)}\times \frac{{\rm SO}(1,m-1)}{{\rm SO}(m-1)}$, using $H_c={\rm SO}(3)\times {\rm SO}(m-1)$, in a two-dimensional universal subspace $\mathfrak{K}_2^{(N)}$ which is common to the all the coset spaces of  $\frac{{\rm SO}(1,3)}{{\rm SO}(3)}\times \frac{{\rm SO}(1,m)}{{\rm SO}(m)}$, including the $m=4$ case. This allows to compute the non-vanishing eigenvalues of a generic $h\in \mathfrak{K}_2$ which are:
\begin{equation}
    \mbox{eigenvalues}(h)=\{ \frac{\kappa_1+\kappa_2}{2},-\frac{\kappa_1+\kappa_2}{2},  \frac{\kappa_1-\kappa_2}{2},-\frac{\kappa_1-\kappa_2}{2},\overbrace{0,\dots, 0}^{m}\}\,,\label{eival8s}
\end{equation}
where $\kappa_\ell$ are real parameters.
The above eigenvalues are compatible with the only orbits \cite{Djokovic}:
\begin{equation}
    [1^{4+m}],\,[2^2,1^m],\,[3,1^{m+1}],\,[3^2,1^{m-2}],\,[5,1^{m-1}]\,,\label{triples}
\end{equation}
which all have non-trivial intersection with the corresponding ${\rm SO}(4,4)$-orbits in (\ref{urbietorbi}).
This motivates our choice of restricting to the $m=4$ manifold for the study of the extremal solutions.\par
The extremal solutions discussed in Sect. \ref{sec:simple} belong, for generic values of $a_1,\,a_2$, to the orbit $[3,1^{5}]$ of ${\rm SO}(4,4)$ and thus the corresponding Noether matrix $Q$ is nilpotent of order three. If $a_1 a_2=0$, the orbit becomes $[2^2,1^4]$ and the same generator is then nilpotent of order two. Below we shall expand on these  two orbits of solutions, leaving a systematic study of solutions belonging to the orbits $[3^2,1^{2}],\,[5,1^{3}]$, and of their supersymmetry properties, to a future work.\par
We conclude that the generating solutions of all the lightlike  geodesics lie within the manifold ${\rm SO}(4,4)/{\rm SO}(1,3)^2$. If  we work in the ${\bf 8}_v$ of ${\rm SO}(4,4)$ instead of the ${\bf 8}_s$, the  orbits $[3,1^5]$ and $[5,1^3]$ are replaced by $[2^4],\,[4^2]$, respectively, so that the maximal order of nilpotency of an element of $\mathfrak{K}^*$in this representation is 4.
Using this property, in Subsection \ref{totalsolution} we give the most general form of the extremal geodesic written in terms of the string moduli $\tilde{G}_{ij},\,C_{ij}$, with boundary conditions $\tilde{G}_{ij}(\rho=0)=\delta_{ij},\,C_{ij}(\rho=0)=0$.\par In the next Subsection we show that, if we only consider the orbits $[2^1,1^4]$ and $[2^4]$ (in the ${\bf 8}_v$), we can restrict ourselves to an even simpler characteristic submanifold $\mathscr{M}_{(N)}=[{\rm SL}(2,\mathbb{R})/{\rm SO}(1,1)]^2$.

\subsubsection{The $[{\rm SL}(2,\mathbb{R})/{\rm SO}(1,1)]^2$ Subspace and Normal Forms for the Orbits $[2^2,1^4],\,[2^4]$}\label{normal}
In this section we construct a characteristic submanifold $\mathscr{M}_{(N)}$ of the Wick-rotated moduli space $\mathscr{M}^*$ which contains representative geodesics of the $[2^2,1^4]$ and the $[3,1^{4}]$ ($[2^4]$ in the ${\bf 8}_v$) orbits. In this way we can relate the abstract and completely general coset construction to the simple Euclidean D1 solutions discussed in section \ref{sec:simple}. The logic presented here was first worked out in detail in \cite{Bergshoeff:2008be} for geodesics on cosets that appear in timelike reductions of supergravity. The general idea is that one truncates the coset to the smallest subspace that generates all geodesics in a certain characteristic subset of all the $G$-orbits by means of the isometry group $G$. This subspace is often, but not always, a simple product of $[{\rm SL}(2,\mathbb{R})/{\rm SO}(1,1)]$ pairs. In light of the discussion in the previous Section we shall restrict ourselves to the Wick-rotated moduli spaces with $m=4$.  

We write $\mathscr{M}^*=G/H^*$ where $G={\rm SO}(4,4)=\exp(\mathfrak{g})$ and $H^*={\rm SO}(1,3)^2=\exp(\mathfrak{H}^*)$.\par

 The isotropy group $H^*$ contains a maximal compact subgroup $H_c=\exp(\mathfrak{H}_1)={\rm SO}(3)^2$, which can be used to simplify the generator $Q\in \mathfrak{K}^*$ of a geodesic. In particular the compact generators in $\mathfrak{K}^*$, which define the axion charges, span the subspace $\mathfrak{H}_{2}$ of $\mathfrak{K}^*$, and transform, under the adjoint action of  $H_c={\rm SO}(3)^2$, in the $({\bf 3},{\bf 1})\oplus ({\bf 1},{\bf 3})$. Similarly the non-compact generators of $\mathfrak{H}^*$ span the subspace $\mathfrak{K}_2$ transforming, under the adjoint action of $H_c$, in the same representation $({\bf 3},{\bf 1})\oplus ({\bf 1},{\bf 3})$ as $\mathfrak{H}_2$. It was shown in Section \ref{INO} that, using ${\rm SO}(3)^2$ transformations, we can always rotate a generic element of $\mathfrak{K}_2$ in a 2-dimensional subspace $\mathfrak{K}_{2}^{(N)}$ (\emph{normal space} of $\mathfrak{K}_2$) generated by two commuting non-compact operators $\mathcal{J}_\ell$, $\ell=1,2$. By the same token, using $H_c$, it is possible to rotate a generic element of $\mathfrak{H}_2$ (describing for instance the compact component of the Noether charge matrix $Q$ of a geodesic) in a 2-dimensional normal subspace $\mathfrak{H}_2^{(N)}$ of $\mathfrak{H}_2$. Let us denote by $\mathcal{K}_\ell$, $\ell=1,2$, a suitable basis of $\mathfrak{H}_2^{(N)}$. As proven in general in \cite{Bergshoeff:2008be} and as we shall show here by direct construction, we can choose $\mathfrak{H}_2^{(N)}$ and $\mathfrak{K}_2^{(N)}$ so that their generators $\mathcal{K}_\ell$ and $\mathcal{J}_\ell$, together with $\mathcal{H}_\ell\equiv [\mathcal{K}_\ell,\,\mathcal{J}_\ell]$, close a characteristic ${\rm SL}(2,\mathbb{R})^2$ subgroup of $G$, and a submanifold
 \begin{equation}
 \mathscr{M}_{(N)}=\left(\frac{{\rm SL}(2,\mathbb{R})}{{\rm SO}(1,1)}\right)^2\subset  \mathscr{M}^*\,,
 \end{equation}
 where the ${\rm SO}(1,1)^2$ at the denominator is generated by $\mathcal{J}_\ell$ and the coset space of $\mathscr{M}_{(N)}$, to be denoted by $\mathfrak{K}_{(N)}$, is generated by $\{\mathcal{H}_\ell,\,\mathcal{K}_\ell\}$. This coset space contains representatives of the $[2^2,1^4]$ and $[2^4]$ (in the ${\bf 8}_v$) orbits and the corresponding geodesics in $ \mathscr{M}_{(N)}$ are easily constructed. Let us define the matrix form of those generators. In the basis of the ${\bf 8}_v$ of ${\rm SO}(4,4)$ used in Section \ref{WRM}, the generators read:
 \begin{align}
 \mathcal{J}_1&=\frac{1}{2}\left({\bf e}_{1,6}-{\bf e}_{2,5}-{\bf e}_{5,2}+{\bf e}_{6,1}\right)\,,\nonumber\\
 \mathcal{J}_2&=\frac{1}{2}\left({\bf e}_{3,8}-{\bf e}_{4,7}-{\bf e}_{7,4}+{\bf e}_{8,3}\right)\,,\nonumber\\
  \mathcal{K}_1&=\frac{1}{2}\left({\bf e}_{1,6}-{\bf e}_{2,5}+{\bf e}_{5,2}-{\bf e}_{6,1}\right)\,,\nonumber\\
 \mathcal{K}_2&=\frac{1}{2}\left({\bf e}_{3,8}-{\bf e}_{4,7}+{\bf e}_{7,4}-{\bf e}_{8,3}\right)\,,\nonumber\\
 \mathcal{H}_1&=\frac{1}{2}\left({\bf e}_{1,1}+{\bf e}_{2,2}-{\bf e}_{5,5}-{\bf e}_{6,6}\right)\,,\nonumber\\
 \mathcal{H}_2&=\frac{1}{2}\left({\bf e}_{3,3}+{\bf e}_{4,4}-{\bf e}_{7,7}-{\bf e}_{8,8}\right)\,,
 \end{align}
 where ${\bf e}_{i,j}$ are matrices with $1$ in the entry $(i,j)$ and $0$ elsewhere.
 Next we define the nilpotent generators $\mathcal{N}_\ell^{(\pm)}$ as follows:
 \begin{equation}
 \mathcal{N}^{(\pm)}_\ell=\mathcal{H}_\ell\mp \mathcal{K}_\ell\,.
 \end{equation}
 These matrices satisfy the relations:
 \begin{equation}
 [\mathcal{J}_\ell,\,\mathcal{N}_{\ell'}^{(\pm)}]=\pm \delta_{\ell\ell'}\,\mathcal{N}_{\ell'}^{(\pm)}\,.
 \end{equation}
 Note that the two sets $\{\mathcal{J}_\ell,\,\mathcal{N}_\ell^{(+)}/\sqrt{2},\,\mathcal{N}_\ell^{(-)}/\sqrt{2}\}$ are standard triples $\{h_\ell,\,e_\ell,\,f_\ell\}$ with nilpotent element $e_\ell$ in the orbit $[2^1,1^4]$, as it can be easily ascertained from the eigenvalues of the neutral elements $h_\ell=\mathcal{J}_\ell$. As shown in the previous section, the most general neutral element $h$ of a standard triple $\{h,e,f\}$ with $e,f\in \mathfrak{K}^*$, modulo an $H_c={\rm SO}(3)^2$ transformation, can be written as $h=\sum_{\ell=1}^2\kappa_\ell h_\ell=\sum_{\ell=1}^2\kappa_\ell \mathcal{J}_\ell$. The eigenvalues of $h$ are:
 \begin{equation}
  \mbox{eigenvalues}(h)=\{\pm \frac{\kappa_1}{2},\pm \frac{\kappa_1}{2},\,\pm \frac{\kappa_2}{2},\,\pm \frac{\kappa_2}{2}\}\,.
\end{equation}
Note the difference between these eigenvalues and those given in (\ref{eival8s}) for $m=4$, which are referred to the same generator in a different, triality-related, representation: the ${\bf 8}_s$.\par
If we try to complete this $h$ into a standard triple $\{h,\,e,\,f\}$, with $e,\,f$ inside the smaller space $\mathfrak{K}_{(N)}={\rm Span}(\mathcal{K}_\ell,\,\mathcal{H}_\ell)$, coset space of $\mathscr{M}_{(N)}$, we see that we only succeed if $\kappa_\ell=0,1,-1$, corresponding to a nilpotent element $e$ in the orbits $[2^2,1^4]$ (for $\kappa_1 \kappa_2=0$) and $[2^4]$ ($\kappa_1 \kappa_2\neq 0$).\footnote{We neglect the trivial case $\kappa_1=\kappa_2=0$.} In both cases this generator would have order of nilpotency 2.\footnote{ If we were working in the ${\bf 8}_s$ we would have the orbit $[3,1^5]$ instead of the $[2^4]$, as explained in the previous section. The corresponding order of nilpotency would then be 3.} Therefore acting by means of $G$ on the lightlike geodesics unfolding in $\mathscr{M}_{(N)}$ one can construct the most general geodesic within the orbits $[2^2,1^4],\,[2^4]$.\par
 The generic nilpotent generator in the coset space $\mathfrak{K}_{(N)}$ has the following form:
 \begin{equation}
 Q=\sum_{\ell=1}^2\,\kappa^{(\pm)}_\ell\,\mathcal{N}^{(\pm)}_\ell\,.\label{Qnil}
 \end{equation}
and has order of nilpotency 2 in the ${\bf 8}_v$. A representative of the orbit $[2^4]$ is obtained when $\kappa^{(\pm)}_1\,\kappa^{(\pm)}_2\neq 0$. Let us illustrate how this orbit splits into suborbits with respect to $H^*$.
Using $H^*$-transformations generated by $h_1,\,h_2$ we can rescale $\kappa^{(\pm)}_1,\,\kappa^{(\pm)}_2$ by a positive factor, so that we can always set $|\kappa^{(\pm)}_\ell|=1$. The inequivalent nilpotent elements in $\mathfrak{K}_{(N)}$ belonging to different $H^*$-orbits can then be reduced to the following four:
\begin{equation}
    \mathcal{N}_1^{(+)}+ \mathcal{N}_2^{(+)},\,  \mathcal{N}_1^{(+)}+ \mathcal{N}_2^{(-)},\, \mathcal{N}_1^{(+)}- \mathcal{N}_2^{(+)},\,  \mathcal{N}_1^{(+)}- \mathcal{N}_2^{(-)}\,,
\end{equation}
and the ${\rm SO}(4,4;\mathbb{C})$-orbit $[2^4]$ split into four  $H^*$-orbits as shown in \cite{Dietrich:2016ojx,Ruggeri:2016dfk}. The signs of $\kappa_\ell^{(\pm)}$ are indeed affected by a transformation of the form $e^{i\pi\,\mathcal{J}_\ell}$ which is in the complexification of $H^*$, while the grading $\pm$ of $\mathcal{N}^{(\pm)}_\ell$ is affected by a transformation of the form $e^{\pi\,\mathcal{K}_\ell}$. Both these transformations are not in $H^*$ and thus different signs of $\kappa_\ell^{(\pm)}$ and different gradings of $\mathcal{N}^{(\pm)}_\ell$ define different $H^*$-orbits\par
The components of $Q$ along the compact generators $\mathcal{K}_\ell$ define the axion charges. Therefore the grading $\pm$ of $\mathcal{N}^{(\pm)}_\ell$ defines the sign of the corresponding axion charge.
\par Let us compute the most general lightlike geodesic in $\mathscr{M}_{(N)}$ passing through the origin.
To this end we define the coset representative in $\mathcal{M}_{(N)}$ in the solvable parametrization, that is we describe the manifold as locally isometric to the solvable group $\exp({\rm Solv})$, where the solvable Lie algebra ${\rm Solv}$ is generated by the matrices $\{\mathcal{H}_\ell,\,\mathcal{T}_\ell\}$, having defined:
\begin{equation}
\mathcal{T}_\ell=(\mathcal{K}_\ell+\mathcal{J}_\ell)\,.
\end{equation}
The coset representative is then defined as follows:
\begin{equation}
L=e^{\sum_\ell {c_\ell}\,T_\ell}\cdot e^{-\sum_\ell {\phi_\ell}\,\mathcal{H}_\ell}\,.
\end{equation}
Next we define the matrix $\mathcal{M}$:
\begin{equation}
\mathcal{M}(\phi)=L(\phi)\eta'L(\phi)^T\,.
\end{equation}
From eq. (\ref{formM}) we can extract from this matrix the matrices $\tilde{G}_{ij}=e^{-\frac{\phi}{2}}\,G_{ij}$ and $C_{ij}$, $G_{ij}$ being the metric of the internal torus in the Einstein frame:
\begin{equation}
    e^{-\frac{\phi}{2}}\,G_{ij}={\rm diag}(e^{-\phi_1},\,e^{-\phi_1},\,e^{-\phi_2},\,e^{-\phi_2})\,\,,\,\,\,\,C_{ij}=\left(
\begin{array}{cccc}
 0 & c_1 & 0 & 0 \\
 -c_1 & 0 & 0 & 0 \\
 0 & 0 & 0 & c_2 \\
 0 & 0 & -c_2 & 0 \\
\end{array}
\right)\,,
\end{equation}
where, using the notation of Section 3,
\begin{equation}
    \phi_1=\frac{\phi+\varphi}{2}-\frac{\psi}{\sqrt{2}}\,\,,\,\,\,\, \phi_2=\frac{\phi+\varphi}{2}+\frac{\psi}{\sqrt{2}}\,.
\end{equation}
The geodesic $\phi(\rho)=\{\phi_\ell(\rho),\,\chi_\ell(\rho)\}$ generated by $Q$, though the origin, is solution to the matrix equation:
\begin{equation}
\mathcal{M}(\phi(\rho))=\mathcal{M}(\phi_0)\,e^{2\rho\,Q}=\eta'\,e^{2\rho\,Q}\,.\label{georigin}
\end{equation}
Solving eq. (\ref{georigin}) we find:
\begin{equation}
c_\ell=\pm\frac{\kappa^{(\pm)}_\ell\,\rho}{H_\ell}\,\,,\,\,\,\,e^{\phi_\ell}=H_\ell\,,
\end{equation}
where $$H_\ell\equiv 1-\kappa^{(\pm)}_\ell\,\rho$$
are harmonic functions. If $\kappa_\ell^{(\pm)}\ge 0$, $H_\ell$ have no poles for $\rho\le 0$ and the solution is regular. The above solution coincides with the one in (\ref{extrsol}) setting $k_\ell=c_{\ell 0}=0$ and $\kappa_\ell^{(\pm)}=a_\ell$. Thus regularity condition selects two out of the four $H^*$-orbits within the complex orbit $[2^4]$.
The grading of the two nilpotent generators is in turn related to the corresponding axion-charge, i.e. to the charges of the Euclidean D1-branes:
\begin{equation}
    q_\ell=\pm \kappa_\ell^{(\pm)}\,.
\end{equation}
Only one choice, that with $q_\ell>0$, defines a supersymmetric configuration. The other, defined by $\kappa_1^{(+)}>0,\,\kappa_1^{(-)}>0$, corresponds to a, extremal, non-supersymmetric, regular solution, in which the two D1 branes have opposite charges.  

\subsection{A Remark on the Regularity Condition for Wormholes}\label{remark}
 Our proof of the non-existence of Euclidean wormholes can be summarized as follows:
 \begin{itemize}
     \item The initial velocity of a timelike geodesic is a compact generator in $\mathfrak{K}^*$ (i.e. an element of $\mathfrak{H}_2$). As discussed in Subsection \ref{normal}, using $H_c$ we can always rotate a generic element of $\mathfrak{H}_2$ into $\mathfrak{H}^{(N)}$, so to be tangent to the normal submanifold $\mathscr{M}_{(N)}$, formally defined in section \ref{normal} and discussed in Section \ref{sec:simple};
\item In Subsection \ref{nowormholes} it is proven that the condition on the maximal length for timelike geodesics in this truncation, for the existence of regular wormhole solutions, is not met.
\end{itemize}
This can also be verified by computing the maximal length $\ell_{{\rm max}}$ on the general timelike geodesic given in Subsection \ref{totalsolution}. This value turns out to be $\ell_{{\rm max}}=\sqrt{2}\pi$ while  regularity of wormhole solutions requires, in three-dimensions, $\ell_{{\rm max}}>2\pi$.\footnote{The corresponding condition in $D$-dimensions is $\ell_{{\rm max}}>2\pi\,\sqrt{\frac{D-1}{2(D-2)}}$. In the truncation discussed in Section 3: $\ell^2_{{\rm max}}=4\pi^2\sum_{i=1}^2\frac{1}{b_i^2}=2 \pi^2$.}\par
We wish here to 
briefly elaborate on the computation of $\ell_{{\rm max}}$ by considering all the inequivalent, totally geodesic ${\rm SL}(2,\mathbb{R})/{\rm SO}(1,1)$ submanifolds of $\mathscr{M}^*$ and the regularity condition (\ref{AHcond}) for the existence of regular wormholes.
The latter condition follows from the requirement that the 
maximal length $\ell_{{\rm max}}$ of timelike geodesics be larger than the actual length of the same curve describing the wormhole solution.  The former quantity $\ell_{{\rm max}}$ is referred to the arc of geodesic comprised between the boundaries of the \emph{physical coordinate patch}, where the scalar fields become singular.  The physical  coordinate  patch is selected by the dimensional reduction of string theory and is defined by the conditions: $$0<G_{ij}<\infty\,,\,\,\,\,\,0<e^\phi<\infty\,.$$
The notion of maximal length is clearly dependent on the coordinate patch and one can find  other local coordinate patches in which the maximal length of a geodesic is larger than in the physical one. A same wormhole solution described in this patch can be regular while being singular when described in terms of the physical fields (coordinates of the physical patch). For example we can consider inequivalent standard triples $\{e,f,h\}$ in $\mathfrak{so}(4,4)$, with $\{e,f\}\subset \mathfrak{K}^*$. The 2-dimensional space $\{e,f\}$ generates a totally geodesic ${\rm SL}(2,\mathbb{R})/{\rm SO}(1,1)$ submanifold of $\mathscr{M}^*$. Restricting to this submanifold and describing the timelike geodesic generated by $e-f$ in the corresponding solvable patch\footnote{The solvable coordinate patch is spanned by a dilatonic scalar and an axionic one, parametrizing the generators $\tilde{h}=\frac{e+f}{\sqrt{2}},\,\tilde{e}=\frac{1}{\sqrt{2}}\left(h-\frac{e-f}{\sqrt{2}}\right)$, respectively, with $[\tilde{h},\,\tilde{e}]=\tilde{e}$. These generators close a solvable Lie algebra. }, one finds $\ell_{{\rm max}}=2\pi/b=\pi\,{\sqrt{d_h}}$ where:
$$ d_h={\rm Tr}({h}\cdot {h})=\sum_{i=1}^\ell {\rm k}_i\,\sum_{m=-s_i}^{s_i} m^2\,,$$
is characteristic of the nilpotent orbit of $e$. If this coordinate patch on the ${\rm SL}(2,\mathbb{R})/{\rm SO}(1,1)$ submanifold were contained in the physical one, the regularity condition would be satisfied for the $[4^2]$ or the $[5,1^3]$ orbit. However this is not the case and along the geodesic within this patch $G_{ij}$ fails to be positive definite. Only the subspaces defined by the triples corresponding to the orbits $[2^2,1^4]$ and $[2^4]$ (or $[3,1^5]$) have their solvable patches embedded in the physical patch on $\mathscr{M}^*$. Both these spaces can be realized within the truncation $\mathscr{M}_{(N)}=[{\rm SL}(2,\mathbb{R})/{\rm SO}(1,1)]^2$ considered in Section 3 and Subsection \ref{normal}. However for these triples $d_h=1$ (for $[2^2,1^4]$) or $d_h=2$ (for $[2^4]$ or $[3,1^5]$) and the regularity condition is not met. Indeed the associated values of $b=2/\sqrt{d_h}$ are $2$ and $\sqrt{2}$, respectively, and the maximal length of timelike geodesics is realized in the latter ${\rm SL}(2,\mathbb{R})/{\rm SO}(1,1)$ submanifold and is $2\pi/b=\sqrt{2}\pi$. This is the same value computed on the general timelike geodesic given in Subsection \ref{totalsolution}.\par
In summary, considering all inequivalent ${\rm SL}(2,\mathbb{R})/{\rm SO}(1,1)$ (totally geodesic) subspaces of $\mathscr{M}^*$ whose (solvable) coordinate patch is contained in the physical patch of the latter, is a valuable approach for assessing the maximal length of timelike geodesics. Each of these 2-dimensional subspaces is defined by a standard triple and is characterized by a value of the $b$-parameter. In the model under consideration only two inequivalent ${\rm SL}(2,\mathbb{R})/{\rm SO}(1,1)$ subspaces satisfy the above requirement and correspond to the partitions  $[2^2,1^4]$ and $[2^4]$ (or $[3,1^5]$) . As pointed out above, both of them are also subspaces of $\mathscr{M}_{(N)}$. \par A similar analysis was implicitly applied to the models considered in \cite{Hertog:2017owm,Ruggeri:2017grz,Katmadas:2018ksp} where  two inequivalent such subspaces exist within the Wick-rotated universal hypermultiplet ${\rm SL}(3,\,\mathbb{R})/{\rm GL}(2,\,\mathbb{R})$, one with $b=2$ and the other with $b=1$. The latter containes the timelike geodesic of maximal length, defining, in that model, a regular wormhole. In this paper we have mathematically formalized and generalized this approach.

\section{Summary and outlook}
Let us summarize the main points of this paper.

\begin{table}[]
\begin{center} 
\begin{tabular}{|l|l|l|} 
\hline
 Orbit & Moduli  & Case  \\ \hline
\small{$Q^4=0$}.&\small{$\tilde{{\bf G}}(\rho)=\left({\bf 1}-2 \rho \,\boldsymbol{\gamma}+2\rho^2\, (\boldsymbol{\gamma}^2+{\bf c}^2)-\frac{4}{3}\,\left((\boldsymbol{\gamma}^2+{\bf c}^2)\cdot \boldsymbol{\gamma} +[\boldsymbol{\gamma},\,\boldsymbol{c}]\cdot \boldsymbol{c}\right)\rho^3\right)^{-1}\,,$ } & \small{Extremal}  \\ 
 &\small{${\bf C}(\rho)=-2 \rho\,\tilde{G}(\rho)\cdot \left({\bf c}- \rho \,[\boldsymbol{\gamma},\,{\bf c}]+\frac{2}{3}\rho^2 \,\left((\boldsymbol{\gamma}^2+{\bf c}^2)\cdot \boldsymbol{c} -[\boldsymbol{\gamma},\,\boldsymbol{c}]\cdot \boldsymbol{\gamma}\right)\right)\,.$}& \\ \hline
\small{$Q^3=0$}.  &\small{$\tilde{ {\bf G}}(\rho)=\left({\bf 1}-2 \rho \,\boldsymbol{\gamma}+2\rho^2\, (\boldsymbol{\gamma}^2+{\bf c}^2) \right)^{-1}\,,$ }& \small{Extremal} \\ 
 &\small{${\bf C}(\rho)=-2 \rho\,\tilde{{\bf G}}(\rho)\cdot \left({\bf c}- \rho \,[\boldsymbol{\gamma},\,{\bf c}]\right).$}& \\ \hline
\small{$Q^2=0$}. &\small{$\tilde{{\bf G}}(\rho)=\left({\bf 1}-2 \rho \,\boldsymbol{\gamma} \right)^{-1}\,,$ }& \small{Extremal}  \\ 
 &\small{${\bf C}(\rho)=-2 \rho\,\tilde{{\bf G}}(\rho)\cdot {\bf c}.$}& \\ \hline
\small{$Q=\sigma_3\otimes \boldsymbol{\gamma}$}, & $\tilde{{\bf G}}(\rho)=\cosh(2\rho \, \boldsymbol{\gamma})+\sinh(2\rho \, \boldsymbol{\gamma})\,,$ & \small{Sub-extremal} \\ 
\small {${\bf c}={\bf 0}$}. & ${\bf C}(\rho)={\bf 0}\,.$ & \\ \hline
\small{$Q=\sigma_1\otimes \boldsymbol{c}$}, & \small{$\tilde{\bf{G}}(\rho)=\cosh(2\rho \, {\bf c})^{-1}\,,$} & \small{Over-extremal}\\ 
\small{$\boldsymbol{\gamma}={\bf 0}$}. & ${\bf C}(\rho)=\sinh(2\rho \, {\bf c})\cdot\cosh(2\rho \, {\bf c})^{-1} \,.$ & \\\hline
\end{tabular}
\end{center}
\caption{\label{T1}{\small The general form of the geodesics on $\mathscr{M}^*$ defined by by a Noether charge matrix $Q=\sigma_3\otimes \boldsymbol{\gamma}+\sigma_1\otimes \boldsymbol{c}$, where $\boldsymbol{\gamma}=\boldsymbol{\gamma}^T,\, \boldsymbol{c}=- \boldsymbol{c}^T$. The elements of the matrix $\tilde{{\bf G}}$ are $e^{-\frac{\phi}{2}}\,G_{ij}$, $\phi$ being the $D=10$ dilaton and $G_{ij}$ the metric moduli of $\mathbb{T}^4$ in the Einstein frame. The matrix elements of ${\bf C}$ are the components of the RR 2-form along the directions of the 4-torus.  }} 
\end{table}

We have classified the instantons in Euclidean $AdS_3\times S^3\times\mathbb{T}^4$ that are carried by the AdS moduli fields dual to the marginal  operators of maximal supersymmetry in the dual CFT. On the supergravity side this corresponds nicely to a classification of geodesics in the moduli space of the Euclidean theory, which we argued boiled down to studying the truncated moduli space
\be
\frac{SO(4,4)}{SO(3,1)\times SO(3,1)}=\frac{SO(4,4)}{SO(4,\mathbb{C})}. 
\ee
We constructed all geodesics and put particular emphasis on the null and timelike cases, see Table \ref{T1}. The null geodesics contain the subgroup of SUSY instantons that lift to Euclidean D1 branes wrapping 2 cycles inside the $\mathbb{T}^4$. It would be interesting to lift all extremal geodesics to 10d and understand their supersymmetry properties.

The timelike geodesics have metrics corresponding to the Giddings-Strominger wormholes \cite{Giddings:1987cg}, but they are not regular in their scalar profile and hence there are no Giddings-Strominger wormholes in this set-up in constrast to the claim in \cite{ArkaniHamed:2007js}. 

We plan to investigate the meaning of the extremal instantons in the dual CFT. The dual CFT is thought to be a 2-dimensional CFT with $(4,4)$ supersymmetries and a central charge proportional to the product $Q_1Q_5$. In the free orbifold point the CFT target space is a large product of $\mathbb{T}^4$ factors divided out by the permutation group \cite{Vafa:1995bm, Witten:1997yu}. Following the procedure of \cite{Tong:2002rq} one could construct the corresponding worldsheet instantons by gauging the sigma-model. To find a correspondence with the supergravity solutions one would hope to find a match between the on-shell actions and the charges. The charges should correspond to charges of the marginal operators dual to the axions.  The latter operators are two-forms
\be
d X^{i}\wedge dX^j\,,
\ee
with $X$ a single copy of the the CFT scalars carrying a vector $SO(4)$-index $i$ under the $SO(4)$-symmetries generated by the internal $\mathbb{T}^4$ torus of the compactification in IIB.
These are closed two-forms that allow for non-trivial topological charges by integration. These should then correspond to the axion charges.

\section*{Acknowledgments}
DA and TVR like to thank the university of Uppsala for hospitality while part of this work was being completed. The work of TVR is supported by the KU Leuven C1 grant ZKD1118C16/16/005.

\appendix


\bibliographystyle{ieeetr}
\bibliography{refs}
\end{document}